\documentclass[twocolumn]{aastex62}

\usepackage{amsmath,bm}
\usepackage{longtable}
\usepackage{natbib}
\usepackage{gensymb}
\usepackage{xcolor}
\usepackage{float}

\usepackage[T1]{fontenc}
\usepackage{newtxtext,newtxmath}
\DeclareRobustCommand{\VAN}[3]{#2}
\let\VANthebibliography\thebibliography
\def\thebibliography{\DeclareRobustCommand{\VAN}[3]{##3}\VANthebibliography}

\defcitealias{trindadefalcao2023a}{Paper 1}
\defcitealias{trindadefalcao2023b}{Paper 2}

\usepackage{graphicx}	
\usepackage{bm}
\usepackage{amsmath}	
\usepackage{booktabs}
\usepackage{gensymb}
\usepackage{soul}
\usepackage{lineno}

\begin{document}


\title[NGC 5728]{Deep \textit{Chandra} Observations of NGC 5728. III: Probing the High-Resolution X-ray Morphology and Multiphase ISM Interactions in the Circumnuclear Region}

\correspondingauthor{Anna Trindade Falcao}
\email{anna.trindade\_falcao@cfa.harvard.edu}

\author{Anna Trindade Falcao}
\affiliation{Harvard-Smithsonian Center for Astrophysics, \\
60 Garden St., Cambridge, MA 02138, USA}

\author{G. Fabbiano}
\affiliation{Harvard-Smithsonian Center for Astrophysics, \\
60 Garden St., Cambridge, MA 02138, USA}

\author{M. Elvis}
\affiliation{Harvard-Smithsonian Center for Astrophysics, \\
60 Garden St., Cambridge, MA 02138, USA}

\author{A. Paggi}
\affiliation{Dipartimento di Fisica, Universita' degli Studi di Torino, via Pietro Giuria 1, \\
I-10125, Torino, Italy}
\affiliation{Istituto Nazionale di Fisica Nucleare, Sezione di Torino, via Pietro Giuria 1, \\
10125, Torino, Italy}

\author{W. P. Maksym}
\affiliation{NASA Marshall Space Flight Center, \\
Huntsville, AL 35812, USA}

\begin{abstract}
We present a detailed imaging analysis of 260 ks of sub-arcsecond resolution \textit{Chandra} Advanced CCD Imaging Spectrometer (ACIS-S) observations of the nearby Seyfert 2 galaxy NGC 5728. Our study focuses on the bright and diffuse soft X-ray emission within the galaxy’s inner $\sim$1~ kpc. By comparing the X-ray emission across different energy bands, we identify localized variations in the absorbing column and emission processes. We observe more X-ray absorption in the direction perpendicular to the bicone, which is co-located with an inner warped CO disk in the galaxy. The innermost region, which shows the strongest excess of hard X-ray emission, is spatially coincident with the CO(2-1) emission from ALMA and dusty spirals we find in a \textit{Hubble Space Telescope} V-H color map. We detect soft extended emission associated with the circumnuclear star-forming ring at $\sim$1~kpc, suggestive of hot gas with $kT=0.44$~keV. We derive measurements for the hot gas mass, $M=7.9\times10^{5}M_{\odot}$, pressure, $p=2.0\times10^{-10}$~dyne~cm$^{-2}$, and cooling times, $\tau_{c}=$193.2 Myr. In the vicinity of the star-forming ring, we detect two X-ray point sources, with soft X-ray spectra, and 0.3-7~keV luminosities $L\sim8\times10^{38}$~erg~s$^{-1}$. These properties suggest X-ray Binaries.
\end{abstract}



\section{Introduction}
\label{sec:introduction}
This is the third paper in our series based on deep \textit{Chandra} observations of NGC 5728 (we refer to \citealt{trindadefalcao2023a, trindadefalcao2023b} as Paper 1 and Paper 2, respectively), a nearby barred spiral galaxy hosting a Compton-Thick (CT, log $N_{H}$=24.2, \citealt{davies2015a}) Seyfert 2 active galactic nucleus (AGN). Due to the CT nature of NGC 5728, the observed X-ray flux of the nuclear source is heavily attenuated, preventing pile-up in ACIS-S observations and reducing the intensity of the point-spread function (PSF) wings. This enables high-resolution exploration of the circumnuclear region down to the smallest subarcsecond radii allowed by \textit{Chandra}’s resolution.

In \citetalias{trindadefalcao2023a}, we examined the spatial and spectral properties of the diffuse extended X-ray emission across the full \textit{Chandra} energy band (0.3–7~keV). The nuclear spectrum ($r<$300~pc) revealed a mixture of low-photoionization and more highly-ionized gas. The extended (300~pc$<r<$1,600~pc) X-ray bicone is dominated by a mix of photoionized and shock-heated gas, suggesting multiple physical processes at play. The diffuse X-ray emission in NGC 5728 is more extended at lower energies, suggesting that optically thick molecular clouds near the nucleus scatter higher-energy AGN photons.

In \citetalias{trindadefalcao2023b}, we reported spatially and spectrally-resolved emission in the Fe K$\alpha$ complex band (5.0-7.5~keV). In the \textit{Chandra} data, these lines are found redward and blueward of the neutral Fe fluorescence line at 6.4~keV, extending to kpc-scales. This suggests an outflow with relativistic velocities of $v\sim$0.1$c$.

In this paper, we focus on high-resolution spectral imaging of the innermost bicone regions and the circumnuclear star-forming ring. We investigate the interaction between the hot and ionized interstellar gas detected with \textit{Chandra} \citepalias{trindadefalcao2023a} and the cold and warm interstellar medium (ISM). The structure of the paper is as follows: Section \ref{sec:observations, data_reduction} outlines the observations and data reduction process, while Section \ref{sec:X-ray_Imaging_Analysis_Results} presents the spectral imaging analysis results. We discuss our findings in Section \ref{sec:discussion}, and our conclusions in Section \ref{sec:conclusions}.

Throughout this paper, we adopt concordance cosmological parameters\footnote{Distances are calculated according to http://www.astro.ucla.edu/$\\sim$wright/ CosmoCalc.html.} of $H_{0}$=70~km~s$^{-1}$~Mpc$^{-1}$, $\Omega_{m,0}$=0.3 and $\Omega_{\Lambda,0}$=0.7. All coordinates are J2000. In all figures, celestial North is up and color scales are logarithmic (unless otherwise noted). For distance evaluation, we use the redshift \textit{z}=0.00932 from NED\footnote{https://ned.ipac.caltech.edu/}, placing NGC 5728 at a distance \textit{D}=41 Mpc with a linear scale of 200 pc arcsec$^{-1}$.

\subsection{The Circumnuclear Region of NGC 5728}
\label{sec:multiwavelength_view}

To provide context for this study, we summarize key findings regarding the circumnuclear region of NGC 5728, particularly the ionization cones and the prominent star-forming ring. Given the galaxy’s proximity, these features are spatially resolved down to scales of $\sim$100~pc or smaller in near-infrared (NIR) and optical wavelengths \citep[e.g.,][]{shimizu2019a, durre2018a, durre2019a, davies2024a}, as well as in X-rays using \textit{Chandra} \citepalias[e.g.,][see Fig. \ref{fig:summary}]{trindadefalcao2023a, trindadefalcao2023b}.

\smallskip
NIR emission-line mapping of the Narrow-Line Region (NLR) using SINFONI \citep[e.g., coronal line region;][]{durre2018a} reveals that the NIR emission lines trace the shape of the optical ionizing bicone, extending to a radius of $r\sim$300~pc. The molecular hydrogen (H$_{2}$) emission aligns with this direction \citep[e.g.,][]{shimizu2019a}, following the large-scale stellar bar that extends over $r\sim11$ kpc, with a position angle of P.A.$\sim$33° \citep{schommer1988a, prada1999a}. \citet{durre2018a} suggest that the H$_{2}$ emission is likely thermally excited by shocks and radiative heating.

Atacama Large Millimeter/submillimeter Array (ALMA) observations of the CO (2-1) emission reveal prominent CO lanes at $r\sim$1 kpc from the nucleus (Fig. \ref{fig:summary}). However, no cold molecular gas is detected in the inner regions where the NIR H$_{2}$ line emission peaks \citep{durre2018a, shimizu2019a}. ALMA also reveals "CO cavities" in a region filled with gas from the soft X-ray bicone \citepalias{trindadefalcao2023a}, likely resulting from the suppression of CO emission by the hard X-ray field from the AGN \citep{shimizu2019a}. Interestingly, the CO emission exhibits a double-peaked structure that straddles the center, perpendicular to the ionization cones \citep[e.g.,][see also Fig. \ref{fig:summary}]{shimizu2019a}. Optical MUSE IFU observations further show that AGN-dominated regions follow the biconical distribution of strong [O~III] emission.

\smallskip
At $r\sim1$ kpc, a star-forming ring surrounds the stellar bar and the ionization cones ($r\sim1.5$ kpc, P.A.$\sim$118°; \citealt{schommer1988a, prada1999a, wilson1993a, mediavilla1995a}). MUSE IFU observations show H~II regions encircling the ionizing bicone in a clear ring-like structure, tracing the ring-shaped emission seen in H$\alpha$ \citep[e.g.,][see Fig. \ref{fig:summary}]{shimizu2019a, durre2018a}.

\section{Observations and Data Reduction}
\label{sec:observations, data_reduction}

\subsection{\textit{Chandra} Observations}
\label{sec:chandra_observations}
NGC 5728 was observed by \textit{Chandra} for a total of 260 ks with ACIS-S, and the details of observations and data reduction are described in \citetalias{trindadefalcao2023a}. The data (Table \ref{tab:observations_log}) were reprocessed following standard procedures, using \texttt{CIAO} (version 4.16\footnote{https://cxc.cfa.harvard.edu/ciao/}) with \texttt{CALDB} 4.11.0 provided by the \textit{Chandra} X-ray Center (CXC), and merged to produce an event file equivalent to a single exposure.

\subsection{Archival Observations}
\label{sec:archival_observations}
\subsubsection{\textit{Hubble Space Telescope (HST)}}
To analyze the extinction in the circumnuclear region of NGC 5728, we have retrieved \textit{HST}/ACS WFC3/F438W (V-band) and WFC3/F160W (H-band) images from the Hubble Legacy Archive\footnote{http://hla.stsci.edu}. These data were acquired as part of Program 13755 (PI: J. Greene, see \citet{shimizu2019a} for details). 

\subsubsection{\textit{SINFONI}}
We also analyzed SINFONI IFU data from the Luminous Local AGN with Matched Analogues (LLAMA) program \citep{davies2015a}. NGC 5728 was observed in February/March 2015 (program ID 093.B-0057, P.I. R. Davies) with the H+K grating (1.4-2.5 $\mu$m) in AO mode, providing a spectral resolution R$\sim$1500 over a $\sim$3$''$$\times$3$''$ field of view (FOV). The final data cube has a pixel scale of 0.05 arcsec/pixel, translating to a spatial resolution of 28 pc over a 660 pc$\times$700 pc FOV. See \citet{shimizu2019a} for details on processing and reduction of these data.

\subsubsection{\textit{ALMA}}
We have used archival ALMA observations of NGC 5728 from program \#2015.1.00086.S (PI: N. Nagar). These observations targeted redshifted CO (2–1) emission ($\nu_{\rm rest}$=230.54 GHz) with an angular resolution of 0.56 arcsec$\times$0.49 arcsec. See \citet{shimizu2019a} for details on processing and reduction.

\begin{table}

\begin{center}
\caption{\textit{Chandra} ACIS-S observations of NGC 5728 used in this work.}
\label{tab:observations_log} 
\begin{tabular}{ccc}
\multicolumn{3}{c}{}\\
\hline
\multicolumn{1}{c}{Observation}
&\multicolumn{1}{c}{Date}
&\multicolumn{1}{c}{Exposure}\\
ID& & Time (ks)\\
\hline
\hline
4077 & 2003-06-27 & 18.73\\

22582 & 2019-12-29 & 49.42\\

22583 & 2020-05-22 & 29.68 \\

23041 & 2020-04-18 & 13.89 \\

23042 & 2021-04-12 & 25.73 \\

23043 & 2020-05-12 & 19.81 \\

23221 & 2020-04-16 & 15.86 \\

23249 & 2020-05-14 & 29.68 \\

23254 & 2020-05-23 & 19.82 \\

25006 & 2021-04-12 & 21.79 \\

25007 & 2021-04-16 & 16.37 \\

\hline
Total& & 260.78\\
\hline
\end{tabular}
\end{center}
\end{table}

\section{High-Resolution X-ray Imaging}
\label{sec:X-ray_Imaging_Analysis_Results}

\begin{figure*}
    \centering
    \includegraphics[width=\textwidth]{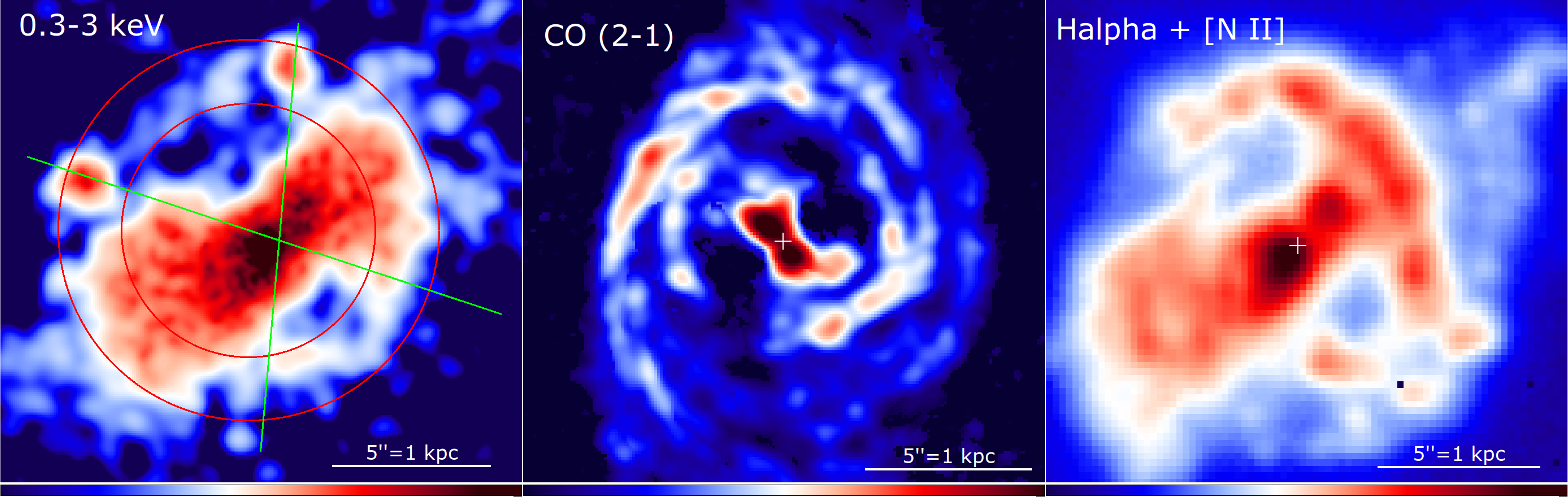}
    \caption{\textit{Left:} 0.3-3 keV \textit{Chandra} adaptive smoothed (9 counts under kernel, 0.5-5 pixels scale, 30 iterations) image of NGC 5728. The different X-ray structures analyzed in the current paper are highlighted in different colors: the innermost soft X-ray bicone (in green), and the circumnuclear star-forming ring (in red). \textit{Central:} ALMA CO (2-1) observations of NGC 5728 show prominent spiral arms of cold molecular gas surrounding the nucleus, and a double-peaked morphology that straddles the center. \textit{Right:} MUSE H$\alpha$+[N~II] map of the central region of NGC 5728 illustrating the prominent star-forming ring observed at $\sim$1 kpc from the central source in the optical. The central and right panels show the location of the \textit{Chandra} X-ray centroid as a white cross.}
    \label{fig:summary}
\end{figure*}

The left panel of Fig. \ref{fig:summary} shows an adaptive smoothed \textit{Chandra} image of NGC 5728 in the soft X-ray band (0.3-3~keV). The central panel shows the same region as seen in the cold molecular gas CO (2-1) line emission, and the right panel, instead, shows the warm and ionized H$\alpha$+ [N~II] gas. In the following subsections, we explore the properties of the X-ray emission across different X-ray-emitting structures, with the goal of developing a more comprehensive view of NGC 5728 through high-resolution \textit{Chandra} imaging. In Section \ref{sec:Surface_Brightness_Distribution_Diffuse_Soft_X-Ray_Emission}, we study the soft X-ray bicone emission (highlighted in green in Fig. \ref{fig:summary}), including a comparison of the emission in different energy bands, and an analysis of the obscuration in different regions within the bicone (Section \ref{sec:Spectral_Mapping_Extended_X-ray_Bicone}). In Section \ref{sec:The_Circumnuclear_Star_Forming _Ring}, we study the X-ray emission from the star-forming ring (highlighted by the red annulus in Fig. \ref{fig:summary}).

\subsection{The X-ray Bicone: Surface Brightness Distribution of the Diffuse X-Ray Emission}
\label{sec:Surface_Brightness_Distribution_Diffuse_Soft_X-Ray_Emission}

\begin{figure*}
    \centering
    \includegraphics[width=\textwidth]{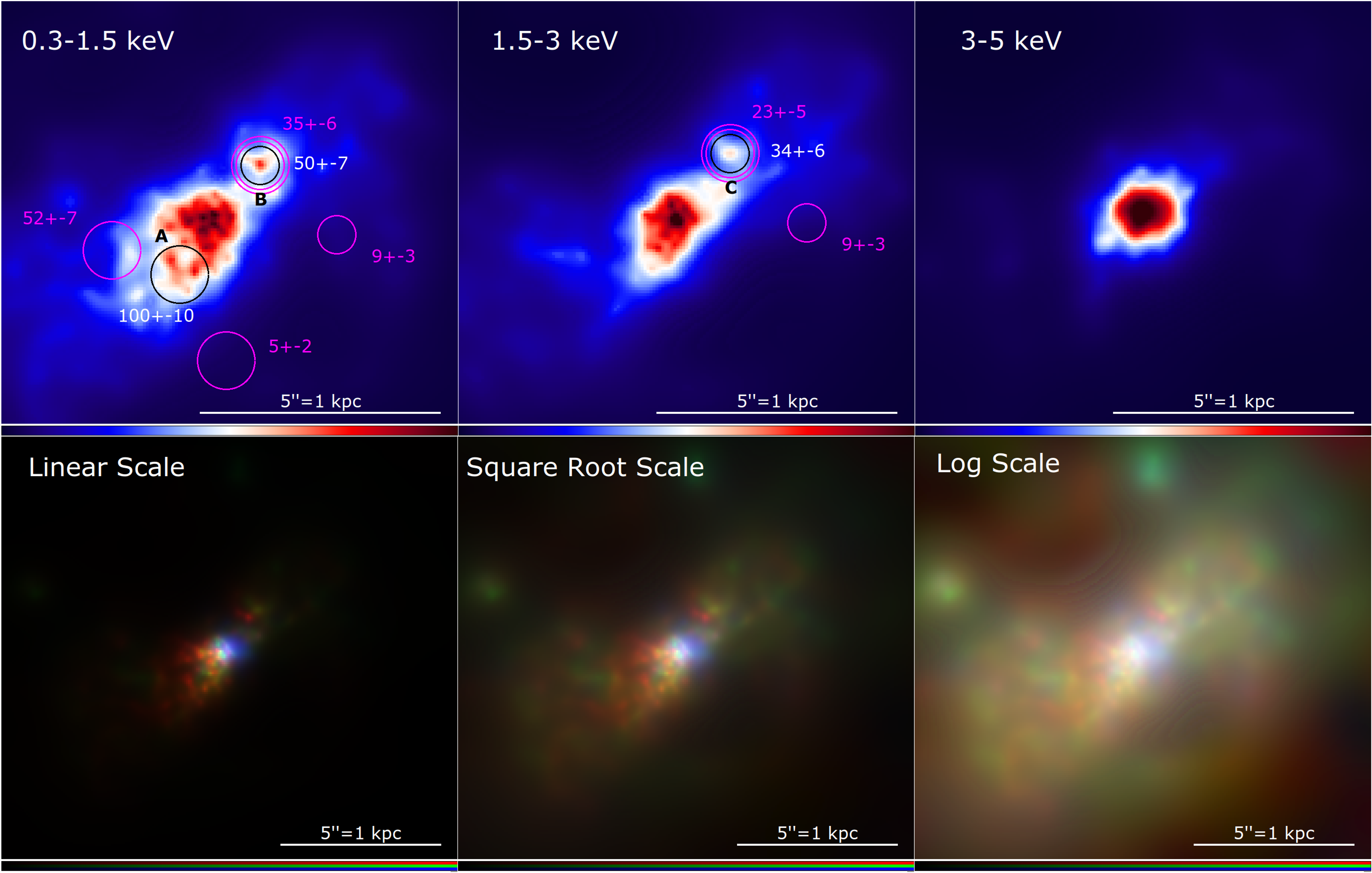}
    \caption{\textit{Top:} \textit{Chandra} ACIS-S images of NGC 5728 in the specified energy bands. These are binned at 1/8 of the native ACIS-S pixel, and are adaptive smoothed with 2-15 pixels scale, 9 counts under kernel, and 30 iterations. Regions of enhanced X-ray emission are indicated in each band. We show the background subtracted number of counts (calculated from the raw data) in each region of interest (black circles, with 1$\sigma$ errors), and compare them to those in equally sized (same area) nearby regions (pink circles and annuli). All images are shown in ASINH scale, for easily comparison between the emission in different bands. \textit{Bottom:} RGB composite images of NGC 5728 shown in different color scales to emphasize different features, linear, square root, and log scale, respectively. Red is the 0.3-1.5 keV image, green is the 1.5-3 keV, and blue is 3-5 keV emission.}
    \label{fig:soft_morphology_rgb}
\end{figure*}

To study the innermost X-ray bicone emission in high-resolution, we create \textit{Chandra} narrow-band images in three distinct energy bands: the soft band, which is dominated by line emission (see \citetalias{trindadefalcao2023a}), further subdivided into 0.3-1.5~keV, and 1.5-3~keV, and the hard continuum band (3-5~keV). These images, shown in the top panels of Fig. \ref{fig:soft_morphology_rgb}, are binned at 1/8 pixel resolution and processed with Gaussian adaptive smoothing (\textit{dmimgadapt}: 2-15 pixel scales, 9 counts under kernel, 30 steps, see \citealt{fabbiano2018b} for previous use of these energy bands and some of the methods in this paper).

Overall, the soft X-ray emission (left and central panels) is extended along the bicone axis, with the SE cone showing brighter and more extended emission than the NW cone. The NW cone appears incomplete - less extended and less prominent than the SE cone - likely due to obscuration of this region by material in the galaxy \citep{wilson1993a, davies2016a}. 

Compared to the soft band ACIS-S images in \citetalias{trindadefalcao2023a}, the adaptively smoothed images in Fig. \ref{fig:soft_morphology_rgb} reveal new X-ray morphological features. This may be attributed to the fact that in \citetalias{trindadefalcao2023a}, our analysis focused on the overall extent of X-ray emission without emphasizing localized surface brightness variations, which are explored here.

0.3-1.5~keV energy band: The very soft emission (Fig. \ref{fig:soft_morphology_rgb}, top-left panel) in the SE bicone is clumpy and has high surface brightness out to $r\sim$400 pc (shown as red features), with fainter diffuse emission extending to $r\sim$600 pc (shown in white). A prominent feature, labeled region "A", shows a $\sim$10$\sigma$ excess and appears as a "hook-like" structure located $r\sim$340 pc from the nucleus. Meanwhile, the NW cone in this band is dominated by fainter diffuse emission out to $r\sim$360 pc, with a $\sim$6$\sigma$ spot of enhanced X-ray emission at $r\sim$260~pc (region "B"). 

1.5-3~keV energy band: In this band (Fig. \ref{fig:soft_morphology_rgb}, top-central panel), the SE bicone has high surface brightness out to $r\sim$230 pc (shown in red), with fainter diffuse emission extending out to $r\sim$400 pc (shown in white). The NW bicone is notably fainter in this band, but emission extends out to $r\sim$360~pc along a similar P.A. A bright spot of enhanced X-ray emission at $r\sim$290~pc from the nucleus (region "C", $\sim$6$\sigma$) is consistent with the location of region "B", within centroid uncertainties ($\sim$0.33$\pm$0.08$''$, or 66$\pm$16 pc, between centroids). 

We show the background subtracted number of counts for each region ("A", "B", and "C") in the top panels of Fig. \ref{fig:soft_morphology_rgb} in white, and compare them to the background subtracted number of counts in identically sized (i.e., same area) nearby regions, in magenta. 

\smallskip
3-5~keV energy band: At higher energies (Fig. \ref{fig:soft_morphology_rgb}, top-right panel), the emission shows a clear peak of surface brightness, although it cannot be explained with a single point source (see radial emission profiles in \citetalias{trindadefalcao2023a}). The emission is marginally extended along the cone axis to $r\sim$300 pc, with lower surface brightness emission extending up to $r\sim$1 kpc in both directions (not shown). These differences between soft and hard X-ray emission in the nuclear region are discussed further in Section \ref{sec:Spectral_Mapping_Extended_X-ray_Bicone}.

\smallskip
Color Composite Images: The bottom panels of Fig. \ref{fig:soft_morphology_rgb} show color composite images (RGB) of the central region of NGC 5728, showing the 0.3-1.5~keV (red), 1.5-3~keV (green), and 3-5~keV (blue) bands. The left-most RGB panel shows the images in linear scale, and emphasizes strong X-ray emission features in each band, particularly along the bicone direction. The middle RGB panel shows these images in square root scale, revealing fainter details of the X-ray emission, mainly along the X-ray bicone. Finally, the right-most RGB panel is shown on log scale, and highlights the faintest X-ray emission in these bands, emphasizing the circumnuclear star-forming ring (predominantly in red and green) and two X-ray point sources located to the north and northeast of the AGN (in green). A central "blue" lane is observed in all RGB panels, indicating either absorption or a relative excess of hard X-ray emission, which is explored in detail below.

\subsubsection{Spectral Mapping of the Extended X-ray Bicone}
\label{sec:Spectral_Mapping_Extended_X-ray_Bicone}

\begin{figure*}
    \centering
    \includegraphics[width=\textwidth]{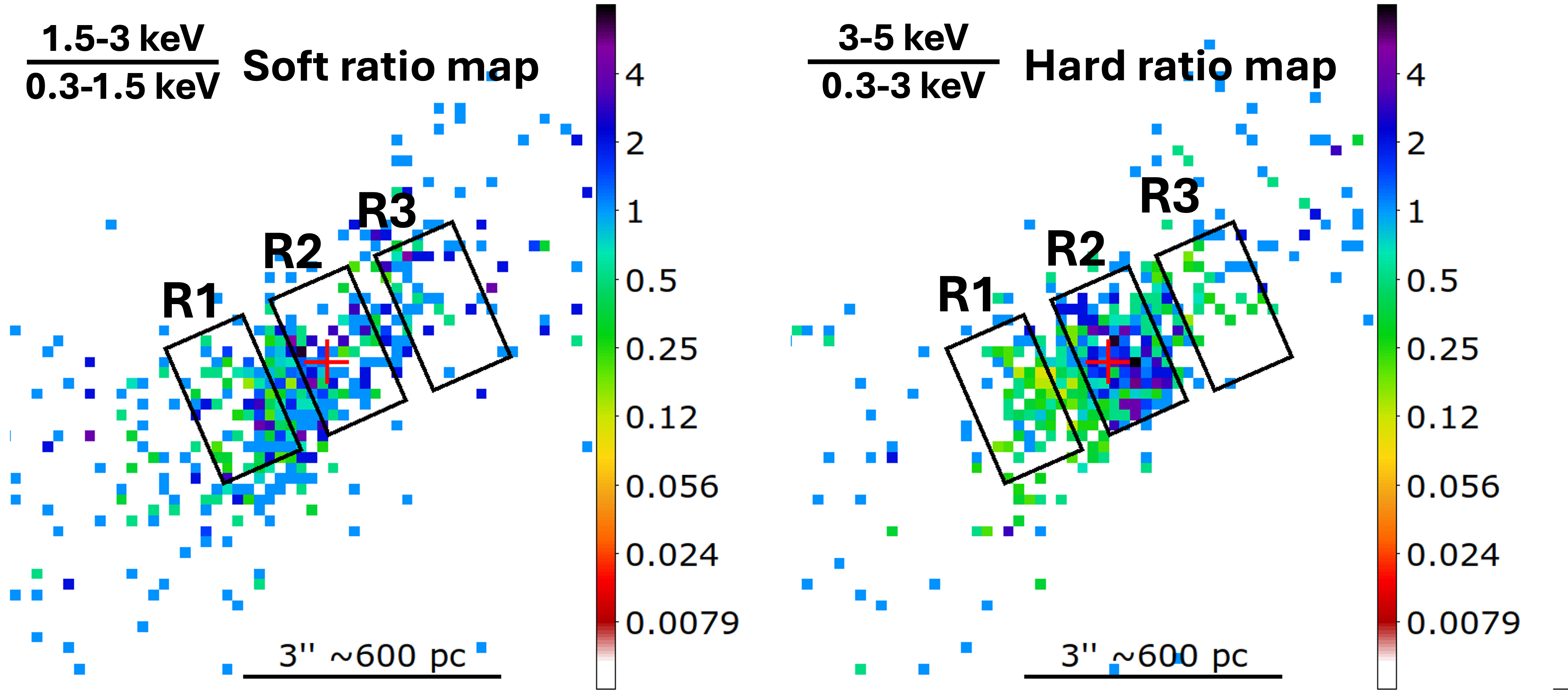}
    \caption{\textit{Left:} Ratio map showing the 1.5-3 keV/0.3-1.5 keV emission ratio in NGC 5728 ("soft ratio map"). \textit{Right:} Ratio map showing the 3-5 keV/0.3-3 keV emission ratio in NGC 5728 ("hard ratio map"). These maps are derived from 1/4 subpixel data, and are shown in log scale. The color scale shows the ratio value per 1/4 ACIS pixel.}
    \label{fig:hardness_ratio}
\end{figure*}

To identify regions of harder X-ray emission or higher obscuration, we create ratio maps showing the 1.5-3~keV/0.3-1.5~keV emission ratio (hereafter the "soft ratio map"), and the 3-5~keV/0.3-3~keV emission ratio (hereafter the "hard ratio map"). These ratio maps, shown in Fig. \ref{fig:hardness_ratio}, were constructed from \textit{Chandra} data binned at 1/4 pixel resolution to improve the signal to noise ratio in each pixel, with no smoothing applied. These maps are sensitive to hydrogen column densities of $10^{22}$~cm$^{-2}$, and $10^{24}$~cm$^{-2}$, respectively, or to excess hard emission ($>$3~keV), reflecting changes in X-ray absorption \citep[e.g.,][]{fabbiano2018b}.

Both the soft and hard ratio maps in Fig. \ref{fig:hardness_ratio} reveal a band of high-value (blue and purple) pixels extending across the nuclear region. This excess of harder emission is more pronounced in the central region (region R2) of the hard ratio map, suggesting a significant and localized excess of hard ($>$3 keV) X-ray emission, potentially due to heavy absorption in this region. We note, however, that the harder emission in this region also includes contributions from the extended hard continuum and Fe K$\alpha$ emission, as previously discussed in \citetalias{trindadefalcao2023a} and \citetalias{trindadefalcao2023b}, and from the point-like nuclear source. Table \ref{tab:hardness_avr_values} shows the average ratio values for regions R1, R2, and R3, along with 1$\sigma$ errors \citep[e.g.,][]{fabbiano2018a}. These regions have sizes 1.7$''\times1.0''$, or 340 pc$\times$200 pc, and probe the X-ray emission at different locations around the nucleus.

Region R1: Located in the SE X-ray bicone, region R1 shows excess softer X-ray emission, characterized by average ratio values $<$1 (Table \ref{tab:hardness_avr_values}). This soft excess likely arises from the elongated morphology of the softer 0.3-3~keV X-ray emission compared to the rounder harder 3-5~keV X-ray emission (Fig. \ref{fig:soft_morphology_rgb}, top panels, and \citetalias{trindadefalcao2023a}). Observed differences in the soft ratio map (Fig. \ref{fig:hardness_ratio}, left) suggest variations in the absorption column, while in the hard ratio map (Fig. \ref{fig:hardness_ratio}, right), these differences imply both differences in absorption and the presence of excess hard X-ray emission.

Region R2: Sampling the innermost nuclear region, region R2 exhibits excess harder X-ray emission (average ratio values $>$1, Table \ref{tab:hardness_avr_values}) in both ratio maps. In the hard ratio map, this excess is oriented along a P.A. $\sim$30$\degree$, roughly perpendicular to the soft X-ray bicone axis.

Region R3: Located in the NW X-ray bicone, region R3 shows an extended, softer (average ratio values $<$1, Table \ref{tab:hardness_avr_values}) morphology in the hard ratio map, similar to region R1. The soft ratio map shows average ratio values $\sim$1.

\begin{figure*}
    \centering
    \includegraphics[width=\textwidth]{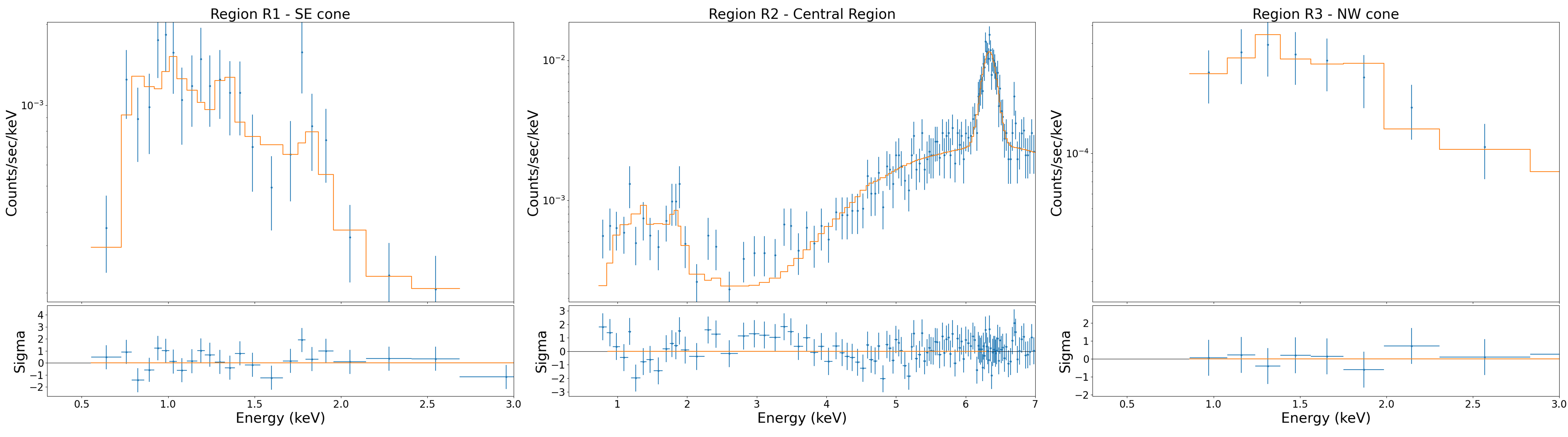}
    \caption{X-ray spectra obtained for the three regions defined in Fig. \ref{fig:hardness_ratio}. The spectra are binned with 15 counts per bin, and fit with an absorbed soft thermal component and a soft power law. For R2, we also include a hard power-law component and a Gaussian for fitting of the hard emission. Best-fit models, and residuals are shown in each panel.}
    \label{fig:bicone_spec}
\end{figure*}

To examine the variations in absorption within the inner region, we use \texttt{CIAO} \textit{specextract}\footnote{https://cxc.cfa.harvard.edu/ciao/ahelp/specextract.html} to extract the spectra from these regions, subtracting the background from a 10$''$ circular region free of point sources. The extracted spectra, binned at 15 counts per bin, are shown in Fig. \ref{fig:bicone_spec}. 

We used \texttt{Sherpa}\footnote{https://cxc.cfa.harvard.edu/sherpa/} to fit the 0.3-3~keV emission in region R1 first, given the improved statistics in this region. The fitting model includes an absorbed soft thermal component and a power-law:

\small{\texttt{xstbabs\footnote{https://cxc.cfa.harvard.edu/sherpa/ahelp/xstbabs.html}.abs\_soft$\times$(xspowerlaw\footnote{https://cxc.cfa.harvard.edu/sherpa/ahelp/xspowerlaw.html}.pl\_soft+xsapec\footnote{https://cxc.cfa.harvard.edu/sherpa/ahelp/xsapec.html}.thermal)}}

\normalsize During the fitting, the soft absorption column density was allowed to vary freely above the Galactic value $N_{H_{\rm galactic}}=7.53\times10^{20}$~cm$^{-2}$ (see \citetalias{trindadefalcao2023a}), while the abundances of the soft thermal component were set to be 1$\times$solar.

\begin{table*}
\centering
\caption{Average ratio values and spectral fitting results for regions R1, R2, and R3, as indicated in Fig. \ref{fig:hardness_ratio}.}
\label{tab:hardness_avr_values} 
\begin{tabular}{cccccccccc}
\multicolumn{3}{c}{}\\
\hline
\multicolumn{1}{c}{Region}
&\multicolumn{1}{c}{Soft Ratio}
&\multicolumn{1}{c}{Hard Ratio}
&\multicolumn{1}{c}{$N_{H}^{\rm soft}$}
&\multicolumn{1}{c}{$\Gamma_{\rm soft}$}
&\multicolumn{1}{c}{$kT$}
&\multicolumn{1}{c}{$N_{H}^{\rm hard}$}
&\multicolumn{1}{c}{$\Gamma_{\rm hard}$}
&\multicolumn{1}{c}{${\rm d.o.f.}$}
&\multicolumn{1}{c}{$\chi^{2}_{\nu}$}\\
&(average count ratio)&(average count ratio)&($10^{21}$~cm$^{-2}$)& & (keV)& ($10^{23}$~cm$^{-2}$)& & &\\
\hline
\hline
R1&0.50$\pm$0.06&0.17$\pm$0.02&4.8$\pm$2.0&1.8$\pm$0.9&0.34$\pm$0.11& & & 18&0.82\\
R2&1.23$\pm$0.10&1.14$\pm$0.08&9.6$\pm$7.4&1.8(f)&0.34(f)&2.2$\pm$1.5&unconstrained& 110& 0.84\\
R3&1.01$\pm$0.20&0.25$\pm$0.04&9.0$\pm$1.6&1.8(f)&0.34(f)& & & 6&0.19\\
\hline
\end{tabular} 
(f): these values were frozen at the best-fit values from the fitting of region R1.
\end{table*}

Region R1 is best-fit with a soft absorbing column $N_{H}$=(4.8$\pm$2.0)$\times$10$^{21}$~cm$^{-2}$, a soft power-law photon index $\Gamma$=1.8$\pm$0.9, and a plasma temperature of $kT$=0.34$\pm$0.11~keV. To obtain better constraints on the absorbing column densities for regions R2, and R3, we fit these spectra with a similar source model, but now freezing the soft power-law photon index and the plasma temperature at the best-fit values from the R1 fit. We leave the soft absorption column and the overall normalization of the source model (soft thermal component normalization/soft power-law component normalization) free to vary during the fitting process.

For region R2, the spectrum at energies $>$3 keV is dominated by emission from the nuclear source, and extended hard and Fe K$\alpha$ emission in the region (see \citetalias{trindadefalcao2023a}, \citetalias{trindadefalcao2023b}, and Fig. \ref{fig:bicone_spec}, center panel). Thus, we fit this spectrum in the 0.3-7~keV band, including an absorbed hard power-law component to account for the nuclear emission, and a redshifted Gaussian to model the neutral Fe K$\alpha$ line at 6.4~keV \citepalias{trindadefalcao2023a}. In \texttt{Sherpa} notation, \\
    \small{\texttt{xstbabs.abs\_soft$\times$(xspowerlaw.pl\_soft+xsapec.thermal)+}\\
\small{\texttt{xstbabs.abs\_hard$\times$(xspowerlaw.pl\_hard+xszgauss\footnote{https://cxc.cfa.harvard.edu/sherpa/ahelp/xszgauss.html}.fe64)}}}

\normalsize The best-fit model yields a soft absorbing column $N_{H}^{\rm soft}$=(9.6$\pm$7.4)$\times10^{21}$~cm$^{-2}$, and a hard absorbing column of $N_{H}^{\rm hard}$=(2.2$\pm$1.5)$\times10^{23}$~cm$^{-2}$, which is associated with the hard nuclear spectral component. The Fe~K$\alpha$ line at 6.4~keV is best fit with a redshifted (\textit{z}=0.00932) Gaussian with $E=(6.40\pm0.01)$~keV, $\sigma=(0.10\pm0.01)$~keV, and normalization=$(2.02\pm0.14)\times10^{-5}$.

Region R3 is best-fit with a soft absorbing column of $N_{H}$=(9.0$\pm$1.6)$\times$10$^{21}$~cm$^{-2}$, consistent with the soft $N_{H}^{\rm soft}$ found in region R2, within uncertainties. 

Best-fit models and residuals for all three regions are shown in Fig. \ref{fig:bicone_spec}, with details in Table \ref{tab:hardness_avr_values}.

\subsection{The Circumnuclear Star-Forming Ring}
\label{sec:The_Circumnuclear_Star_Forming _Ring}

\begin{figure*}
    \centering
    \includegraphics[width=\textwidth]{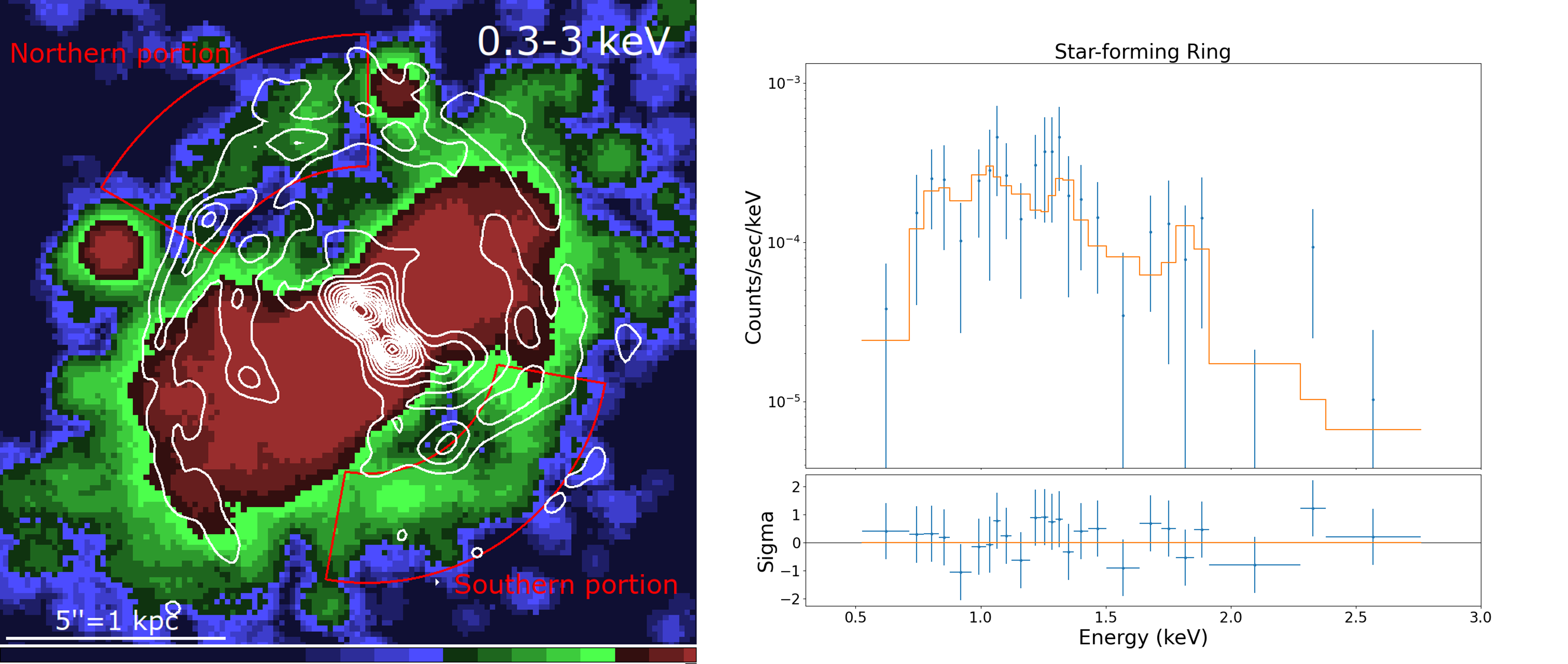}
    \caption{\textit{Left:} Smoothed \textit{Chandra} 0.3-3~keV image of NGC 5728 binned at 1/4 of the native ACIS-S pixel. Contours from CO(2–1) emission from ALMA \citep{ramakrishnan2019a} are shown in white. Red sectors represent the regions used for spectrum extraction. \textit{Right:} Merged X-ray spectrum of the star-forming ring in NGC 5728. The extraction regions used are shown in the left panel, and the data is binned at 5 counts per bin. We show the best-fit spectral model in orange, and residuals on the bottom panel.}
    \label{fig:chandra_ring}
\end{figure*}

With \textit{Chandra}, we detect X-ray emission from the prominent circumnuclear star-forming ring located at a radius $r\sim$1~kpc ($r\sim$5$''$) from the nuclear AGN \citep[e.g.,][]{shimizu2019a}. This ring is highlighted in red in the left panel of Fig. \ref{fig:summary}. Fig. \ref{fig:chandra_ring} shows the smoothed 0.3-3~keV \textit{Chandra} X-ray image of this region, binned at 1/4 pixel resolution.

The southern portion of the  star-forming ring is brighter than the northern portion, with an integrated surface brightness of 7.1$\pm$0.7 counts/arcsec$^{2}$, within the $3''<r<5.5''$ region.  In contrast, the northern portion of the ring has 3.8$\pm$0.5 counts/arcsec$^{2}$ within the $4''<r<7''$ region, 3.8$\sigma$ below the value for the southern portion.

In the left panel of Fig. \ref{fig:chandra_ring} we overlay white contours of CO (2-1) emission from ALMA \citep{ramakrishnan2019a}. The CO emission traces the inner edges of the X-ray-emitting gas along the ring, in addition to a transverse dust lane crossing the nuclear region. This lane connects the two ends of the ring, creating a "CO cavity" filled with X-ray emitting gas, the X-ray bicone.

We use \texttt{CIAO} \textit{specextract} to extract the spectra from the ring regions shown in the left panel of Fig. \ref{fig:chandra_ring}. The merged X-ray spectrum of the star-forming ring, binned at 5 counts per bin, is shown in the right panel. We note that, given the limited number of counts per bin, we cannot assume a Gaussian distribution; instead, the standard deviation is derived following \citet{gehrels1986a}, using the \textit{chi2gehrels} model\footnote{https://cxc.cfa.harvard.edu/sherpa/ahelp/chi2gehrels.html}.

We fit the 0.3-3~keV spectral data with an absorbed single-temperature thermal model, \\
\texttt{(xstbabs.abs\_soft$\times$xsapec.thermal)}.\\
The absorption column density was left free to vary above the Galactic value, and the abundance of the thermal component was set to 1$\times$solar. Although the gas could contain multiple temperature components, the low signal-to-noise ratio (S/N) of the data prevents us from constraining a multi-temperature plasma. Our best-fit model ($\chi^{2}_{\nu}=0.46$) yields a temperature of $kT$=0.44$\pm$0.17 keV, and an absorbing column $N_{H}=(6.0\pm2.1)\times10^{21}$~cm$^{-2}$.

Using \texttt{CIAO} \textit{calc\_energy\_flux}\footnote{https://cxc.cfa.harvard.edu/sherpa/ahelp/calc\_energy\_flux.html}, we estimate the observed X-ray flux in the soft 0.3-3~keV energy band to be ${f}^{\mathrm{model}}_{0.3-3.0\:\mathrm{keV}}=1.93\times10^{-15}$~erg~cm$^{-2}$~s$^{-1}$. This corresponds to an X-ray luminosity of ${L}^{\mathrm{model}}_{0.3-3.0\:\mathrm{keV}}=3.9\times10^{38}$~erg~s$^{-1}$.

\subsubsection{Two Luminous X-ray Point Sources}
\label{sec:Two_Luminous_X-ray_Point_Sources}

Two bright X-ray point sources, NGC~5728~X-1 and NGC~5728~X-2, were detected near the star-forming ring (Figs. \ref{fig:soft_morphology_rgb} and \ref{fig:chandra_ring}), primarily emitting in the soft ($<$3~keV) X-ray band \citepalias[see also][]{trindadefalcao2023a}. We fit their spectra using three models: a thermal model, a power-law model, and an accretion disk model. The estimated X-ray luminosities for both sources range between $L_{\rm 0.3-7~keV}=5.0-8.2\times10^{38}$ erg s$^{-1}$, consistent with High Mass X-ray Binaries (HMXBs), suggesting these objects may significantly contribute to the observed X-ray luminosity in this band. Further details on the analysis and spectral fitting results are provided in the Appendix \ref{sec:appendix}.

\section{Discussion}
\label{sec:discussion}

In this study, we focused primarily on the properties of the soft (0.3-3~keV) X-ray emission within the extended soft X-ray bicone and the circumnuclear star-forming ring of NGC 5728.

We analyzed narrow-band \textit{Chandra} images of this galaxy, and identified localized regions of enhanced soft X-ray emission within the bicone (Fig. \ref{fig:soft_morphology_rgb}, Section \ref{sec:Surface_Brightness_Distribution_Diffuse_Soft_X-Ray_Emission}). Additionally, image ratio maps were created to identify regions with excess harder X-ray emission (Fig. \ref{fig:hardness_ratio}, Section \ref{sec:Spectral_Mapping_Extended_X-ray_Bicone}). Our spectral analysis revealed a range of column densities within the innermost regions of the bicone, from $N_{H}$=4.8$\times$10$^{21}$~cm$^{-2}$ to 2.2$\times10^{23}$~cm$^{-2}$ (Table \ref{tab:hardness_avr_values}).

At a radius of $r\sim$1 kpc from the nuclear AGN, we detected a circumnuclear star-forming ring in X-rays with \textit{Chandra} (Fig. \ref{fig:chandra_ring}). We modeled the X-ray emission in the ring with a collisionally-ionized diffuse component, and obtained constraints on the temperature, k$T$=0.44$\pm$0.17~keV, and absorbing column, $N_{H}=(6.0\pm2.1)\times10^{21}$~cm$^{-2}$, of the hot gas in this region (Section \ref{sec:The_Circumnuclear_Star_Forming _Ring}). Additionally, we analyzed the X-ray emission of two point-like sources located near the star-forming ring, first reported in \citetalias{trindadefalcao2023a}. These sources appear brighter in the 0.3-3~keV band, with X-ray luminosities ranging from $L_{\rm 0.3-7~keV}=5.0-8.2\times10^{38}$~erg~s$^{-1}$ (Appendix \ref{sec:appendix}, Table \ref{tab:ps}). 

Below we discuss the implications of our results, focusing on:

1) Bicone Obscuration (Section \ref{sec:A Multi-wavelength View of the Bicone Obscuration});

2) Connection Between Soft X-ray Bicone Emission and the Multiphase ISM (Section \ref{sec:Bicone H2, CO and X-ray emission});

3) Properties of the hot X-ray emitting gas and point-like sources in the Star-Forming Ring (Section \ref{sec:The Star-forming Ring}). 

\subsection{A Multi-wavelength View of the Bicone Obscuration}
\label{sec:A Multi-wavelength View of the Bicone Obscuration}

\subsubsection{Extinction}
\label{sec:extinction}
\begin{figure}
    \centering
    \includegraphics[width=.5\textwidth]{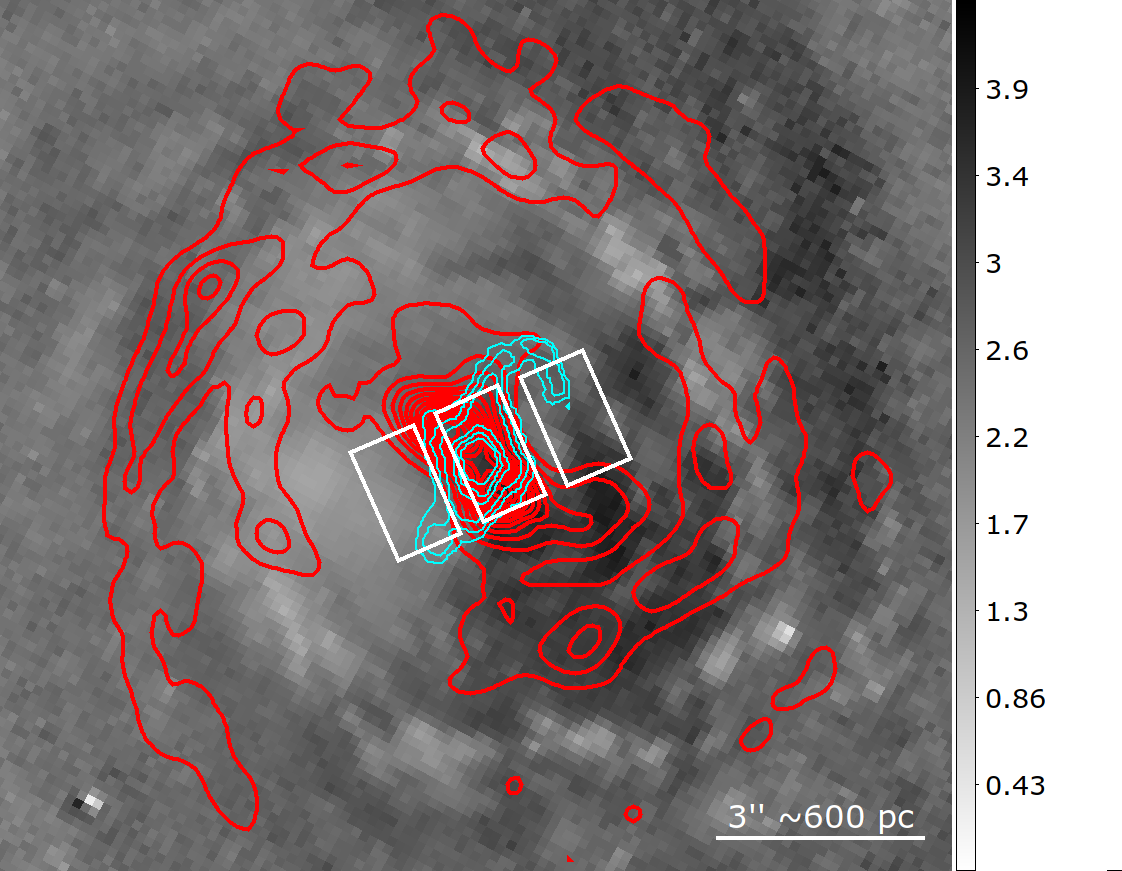}
    \caption{V-H color map of NGC 5728 constructed using archival \textit{HST} F438W (V-band) and F160W (H-band) images. Regions with high extinction have large V-H values, and appear darker in the map. The color scale shows the V-H (mag) values. We also show the position of the 3 regions as in Fig. \ref{fig:hardness_ratio}, the CO (2-1) contours from ALMA in red, and the H$_{2}$ contours from SINFONI in cyan.}
    \label{fig:vh}
\end{figure}

To analyze the overall extinction in the circumnuclear region of NGC 5728, we first constructed a V-H color map using archival \textit{HST} F160W and F438W images (Section \ref{sec:archival_observations}). This color map, shown in Fig. \ref{fig:vh}, reveals the distribution of dust and highlights spiral features of higher extinction near the nucleus.

For comparison, we overlaid the three regions (R1, R2, R3) defined in Section \ref{sec:Spectral_Mapping_Extended_X-ray_Bicone} and Fig. \ref{fig:hardness_ratio}  as white rectangles. In addition, we plotted the CO (2-1) contours from ALMA (in red), and the H$_{2}$ (1-0) S(1) contours from SINFONI (in cyan). The color scale is linear and shows the V-H values (in mag).

The V-H values derived from the color map can be used to estimate the extinction in each rectangular region. For this, we rely on the stellar population synthesis results from \citet{burtscher2021a}, which estimate the age of the nuclear stellar population in NGC 5728 to be log(age/yr)$>9.5$, indicating a generally old stellar population in the inner regions. Using the extinction curve from \citet{cardelli1989a}, the observed E$_{\rm V-H_{\rm map}}$ values translate into estimates for E$_{\rm B-V_{\rm map}}$ and A$_{\rm V_{\rm map}}$, as summarized in Table \ref{tab:reddening}.

\begin{table}
\begin{center}
\caption{Range in extinction derived for regions R1, R2, and R3, from the \textit{HST} V-H map and from the spectral analysis in Section \ref{sec:Spectral_Mapping_Extended_X-ray_Bicone}.}
\label{tab:reddening} 
\begin{tabular}{ccccc}
\multicolumn{5}{c}{}\\
\hline
\multicolumn{1}{c}{Region}
&\multicolumn{1}{c}{V-H$_{\rm map}$}
&\multicolumn{1}{c}{E$_{\rm B-V_{\rm map}}$}
&\multicolumn{1}{c}{A$_{\rm V_{\rm map}}$}
&\multicolumn{1}{c}{A$_{\rm V_{\rm spec}}$}\\
& (mag)&(mag) & (mag)& (mag)\\
\hline
\hline
R1&1.8-2.7&0.1-0.5&0.4-1.5&1.4-3.4 \\
R2&2.7-3.9&0.5-1.0&1.5-3.1&1.1-8.5\\
R3&2.1-3.7&0.2-0.9&0.8-2.8&3.7-5.3\\
\hline
\end{tabular}
\end{center}
\end{table} 

By applying the relation $N_{H}$/$A_{\rm V}\sim$ 2.0$\times$10$^{21}$~cm$^{-2}$ \citep[e.g.,][]{ryter1996a}, we estimate extinction values for each region using the results of the spectral analysis in Section \ref{sec:Spectral_Mapping_Extended_X-ray_Bicone}. As shown in Table \ref{tab:reddening}, the extinction values obtained from the V-H map, A$_{\rm V_{\rm map}}$, are consistent with the extinction values estimated from the results of the X-ray spectral analysis, A$_{\rm V_{\rm spec}}$, within uncertainties.  

\subsubsection{Line-of-Sight Nuclear Obscuration}
\label{sec:lineofsight_obs}
The hard ratio map in Fig. \ref{fig:hardness_ratio} does not show evidence of a localized excess consistent with a point-like obscured source, as observed in ESO 428-G014 \citep{fabbiano2018b}. Instead, a general band of high ratios - indicating an excess of hard photons - is visible perpendicular to the bicone direction. This high ratio band aligns with the central CO structure (Fig. \ref{fig:vh}), and the extent of both high X-ray band ratios and the CO lane are similar. 

In NGC 5728, \citet{durre2018a} estimated a cold gas column density of 420~$M_{\odot}$~pc$^{-2}$, equivalent to $N_{H} \sim 4 \times 10^{22}$~cm$^{-2}$, based on Pa$\alpha$ and Br$\gamma$ flux maps. This value suggests a lower line-of-sight gas column density compared to the hard $N_{H}^{\rm hard} = (2.2 \pm 1.5) \times 10^{23}$~cm$^{-2}$ derived from the X-ray spectral analysis (Table \ref{tab:hardness_avr_values}).

It is possible that the CO arm-like structure observed in the nuclear region (Fig. \ref{fig:vh}) passes in front of the X-ray emission, obscuring it, as seen in ESO 428-G014 \citep{feruglio2020a}. However, the V-H map (Fig. \ref{fig:vh}) does not show enhanced absorption in the R2 region. Additionally, the soft $N_{H}^{\rm soft}$ value derived in Section \ref{sec:Spectral_Mapping_Extended_X-ray_Bicone} for R2 is comparable to those in R1 and R3. This suggests that any obscuration by the CO lane contributing to the hard $N_{H}^{\rm hard}$ derived for R2 must occur internal to most of the stellar bulge. The emission lines used by \citet{durre2018a} to probe nuclear obscuration likely originate from regions closer to the nucleus than the bulge stars observed in the V-H map.

Other studies of NGC 5728 using \textit{NuSTAR} data have reported even higher line-of-sight gas columns, with $N_{H} > 10^{24}$~cm$^{-2}$ \citep[e.g.,][]{marchesi2018a, tanimoto2022a}. The discrepancy between these measurements may arise from the different energy ranges probed by the studies. \textit{NuSTAR}'s higher energy range (3-79~keV) allows for better detection of the harder, unabsorbed power-law component, facilitating more accurate measurements of $N_{H}$. Additionally, the higher energy coverage of \textit{NuSTAR} (above 3~keV) reduces contamination from extended emission, which can affect instruments such as \textit{Chandra}.

\subsection{The Interplay Between the Molecular ISM and X-ray Emission in the Bicone}
\label{sec:Bicone H2, CO and X-ray emission}

In Fig. \ref{fig:vh}, we compare the distribution of cold molecular gas (CO, shown in red) and warm molecular gas (H$_{2}$, shown in cyan) with the X-ray obscuration observed in the inner regions of NGC 5728 (white rectangles). In this subsection, we analyze the interplay between these different gas phases within the ISM of NGC 5728.

\begin{figure*}
    \centering
    \includegraphics[width=\textwidth]{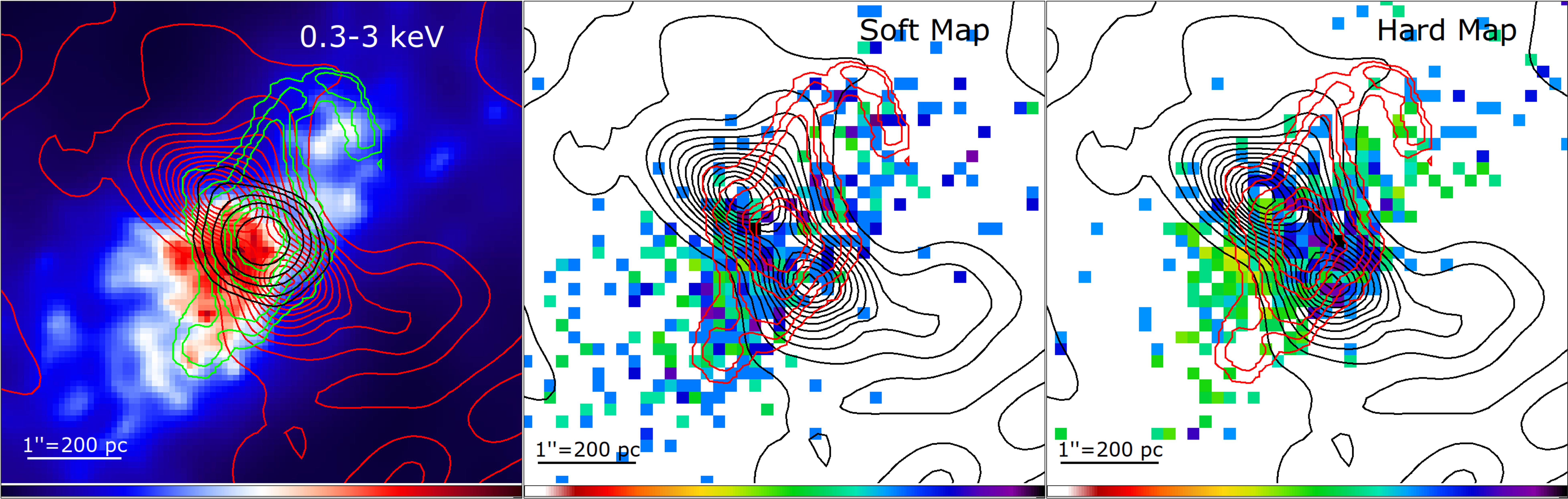}
    \caption{\textit{Left:} \textit{Chandra} ACIS-S images of NGC 5728 in the 0.3-3~keV band. This image is binned at 1/8 of the native ACIS-S pixel, and is adaptive smoothed with 2-15 pixels scale, 9 counts under kernel, and 30 iterations. Overlaid are the 1.3 mm contours from ALMA (in black), CO contours from ALMA (in red), and warm molecular H$_{2}$ contours from SINFONI (in green). \textit{Center and Right:} Soft and hard ratio maps, respectively, as in Fig. \ref{fig:hardness_ratio}, overlaid with H$_{2}$ contours from SINFONI in red and CO contours from ALMA in black.}
    \label{fig:h2_analysis}
\end{figure*}

Fig. \ref{fig:h2_analysis} (left) shows an adaptively smoothed \textit{Chandra} image of NGC 5728 in the 0.3-3~keV band, overlaid with contours of H$_{2}$ from SINFONI (in green), 1.3 mm continuum from ALMA (in black), and CO(2-1) from ALMA (in red). The 1.3 mm continuum is marginally extended ($r\sim$200 pc) in the cross-cone direction, and peaks at the X-ray nucleus. In contrast, the CO emission shows a double-peaked morphology, straddling the center and oriented perpendicular to the X-ray bicone, as previously mentioned. Meanwhile, the H$_{2}$ contours are extended along the X-ray bicone, reaching $r\sim$400~pc into the NW cone and $r\sim$320~pc into the SE cone. These H$_{2}$ contours are co-spatial with the brightest regions of soft X-ray emission (shown in white and red).

The distribution of the cold and warm molecular gas, along with the X-ray emission in the circumnuclear region, is consistent with findings from \citetalias{trindadefalcao2023a}. In that study, a hard ($>$3 keV) X-ray component (Fig. \ref{fig:soft_morphology_rgb}, top right panel) was observed extending along the soft X-ray bicone, suggesting reflection and fluorescence off molecular clouds. Such extended hard X-ray features are common in CT AGNs observed with \textit{Chandra}, as well as their connection to molecular clouds in the NLR \citep{fabbiano2024a}.

As discussed in \citetalias{trindadefalcao2023a} and shown in the ratio image maps in Fig. \ref{fig:hardness_ratio}, we observe regions of harder X-ray emission in the soft ratio map (1.5-3~keV/0.3-1.5~keV) along the direction of the X-ray bicone. Fig. \ref{fig:h2_analysis} (center and right panels) shows that these regions of harder emission are co-located with warm molecular H$_{2}$ emission (shown in red). For comparison, the CO contours (in black) are also shown, highlighting the relative positioning of cold molecular gas.

\begin{figure}
    \centering
    \includegraphics[width=.5\textwidth]{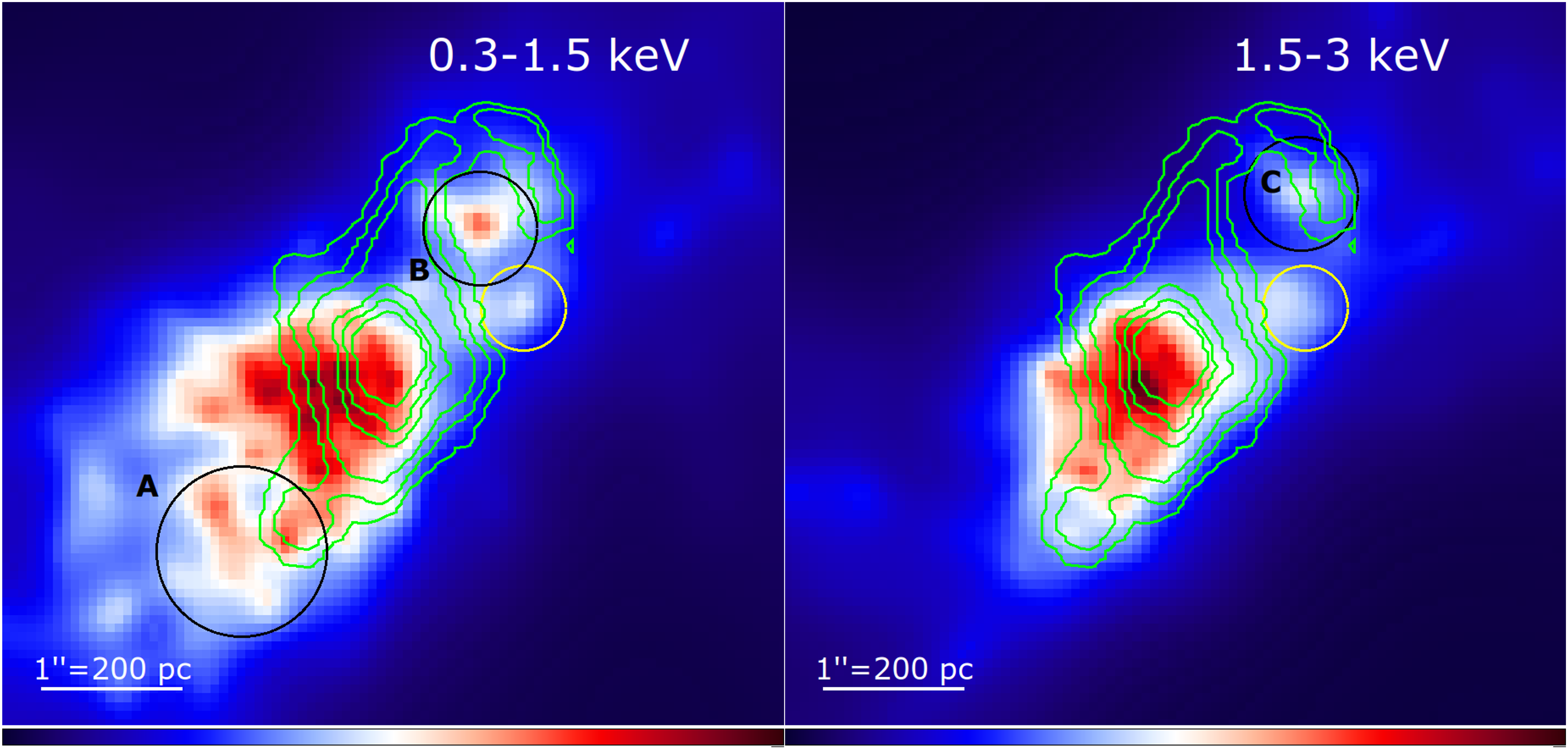}
    \caption{\textit{Chandra} ACIS-S images of NGC 5728 in the specified energy bands. These are binned at 1/8 of the native ACIS-S pixel, and are adaptive smoothed with 2-15 pixels scale, 9 counts under kernel, and 30 iterations. Overlaid in green are the H$_{2}$ contours from SINFONI. We also mark the regions of enhanced X-ray emission A, B, and C, as in Fig. \ref{fig:soft_morphology_rgb}, and an additional region of enhanced emission as a yellow circle.}
    \label{fig:h2_analysis_morp}
\end{figure}

\subsubsection{Bicone H$_{2}$ and X-ray Emission}

To explore the connection between the regions of excess harder X-ray emission observed in the ratio maps and the H$_{2}$ morphology, Fig. \ref{fig:h2_analysis_morp} shows two \textit{Chandra} images of NGC 5728 in the soft X-ray band (as in Fig. \ref{fig:soft_morphology_rgb}, center and right), overlaid with H$_{2}$ contours from SINFONI (green). The regions of enhanced X-ray emission, labeled A, B, and C, which were previously identified in Section \ref{sec:Surface_Brightness_Distribution_Diffuse_Soft_X-Ray_Emission}, are also marked, for comparison.

Region "A" (100$\pm$10 excess counts) is spatially-coincident with the outskirts of the H$_{2}$ contours southeast of the nucleus. Region "B" (50$\pm$7 excess counts) is located "in-between" H$_{2}$ contours northwest of the nucleus. This region may represent the point of impact where the radio jet interacts with the ISM \citep[e.g.,][]{durre2018a}, leading to localized heating and enhanced X-ray emission.

In the 1.5-3~keV band, region "C" (34$\pm$6 excess counts) is spatially-coincident with the "hook" H$_{2}$ feature observed northwest of the central source. The close alignment between the X-ray and H$_{2}$ features in this region suggests a direct connection, potentially driven by the interaction of the nuclear radio jet with the ISM.

In both X-ray bands, an additional region of enhanced X-ray emission is observed between the H$_{2}$ "hook" and the H$_{2}$ contours west of the nucleus (highlighted by yellow circles in Fig. \ref{fig:h2_analysis_morp}). This region may also be attributed to the impact of the radio jet on the ISM, leading to increased excitation in the region \citep{durre2018a}.

\subsubsection{H$_{2}$ Excitation}

There are three primary excitation mechanisms for H$_{2}$: 

(1) Thermal Processes: shocks from supernovae (SNe), star-formation winds or AGN outflows \citep{hollenbach1999a}; or X-rays from the AGN, which heat and irradiate dense gas \citep{maloney1996a}.

(2) Non-Thermal Processes: UV photons (fluorescence) from AGN continuum emission or star formation \citep{black1987a}. This process involves UV photons with wavelengths in the range 912\AA$<\lambda<$1500\AA, which are absorbed by H$_{2}$ in the Lyman and Werner bands, exciting the two next electronic levels. It also populates higher H$_{2}$ ro-vibrational levels that are not typically populated by collisions, and is often driven by sources like OB stars or strong AGN emission.  

While all three excitation mechanisms may occur simultaneously, H$_{2}$ line fluxes can help constrain the dominant excitation mechanism and the relative contributions of each process \citep{busch2017a}. For example, the line ratio of H$_{2}$ 2-1 S(1) to H$_{2}$ 1-0 S(1) can distinguish between thermal excitation (from shocks or X-ray heating) and non-thermal excitation (by soft UV photons from star-formation). Ratios of $\sim$0.1-0.2 indicate thermal processes, while ratios $\sim$0.55 suggest non-thermal excitation \citep[e.g.,][]{riffel2014a}. Additionally, ratios such as H$_{2}$ 1-0 S(2)/H$_{2}$ 1-0 S(0) and H$_{2}$ 1-0 S(3)/H$_{2}$ 1-0 S(1) can further differentiate between shocks, thermal UV, and X-ray excitation \citep[e.g., diagnostic diagram by][]{mouri1994a}.

In NGC 5728, the H$_{2}$ \textit{K}-band line fluxes were measured at different locations by \citet{durre2018a} (Fig. 19 therein). By computing the vibrational and rotational temperatures of H$_{2}$ at these locations and plotting the results on a \citet{mouri1994a} excitation diagram, they found that the H$_{2}$ gas in NGC 5728 is primarily excited by thermal processes, such as shocks and/or X-ray heating. The H$_{2}$ 2-1 S(3) (2073.5 nm) line, which is diagnostic of X-ray excitation, was detected at all locations analyzed by \citet{durre2018a}. However, the low line fluxes detected in the innermost bicone regions suggest that shocks are the dominant excitation mechanism in this region, with X-rays likely playing a minor role \citep[e.g.,][for NGC 1275]{krabbe2000a}. This conclusion is consistent with the X-ray spectral analysis from \citetalias{trindadefalcao2023a}, which found that the soft X-ray bicone emission is dominated by a mix of photoionized and collisionally-ionized gas, further supporting the role of shocks and thermal excitation processes. 

\subsection{The Star-forming Ring}
\label{sec:The Star-forming Ring}

\textit{Chandra}/ACIS-S X-ray imaging of NGC 5728 reveals hot ISM within a star-forming ring at $r\sim$600~pc (Section \ref{sec:The_Circumnuclear_Star_Forming _Ring}, Fig. \ref{fig:chandra_ring}). In this section, we compare the soft X-ray emission in this region with archival observations from \textit{HST} F160W, \textit{HST} F438W, and MUSE H$\alpha$+[N~II]. 

\begin{figure*}
    \centering
    \includegraphics[width=\textwidth]{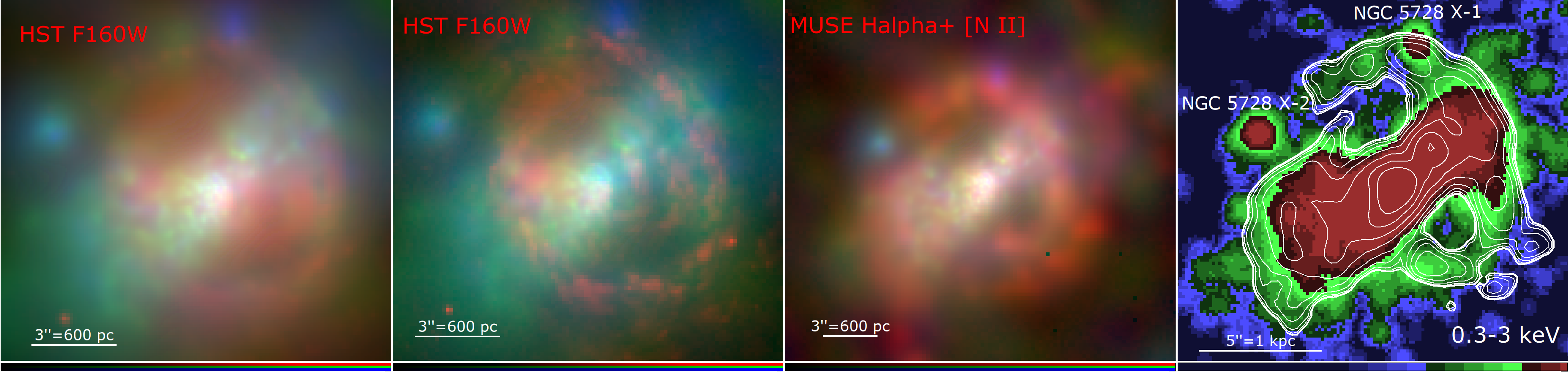}
    \caption{\textit{First:} RGB composite image of NGC 5728. \textit{HST} F160W image in shown in red, 0.3-1.5 keV soft X-ray emission is shown in green, and 1.5-3 keV soft X-ray emission is shown in blue; \textit{Second:} As in the first panel, but \textit{HST} F438W image shown in red; \textit{Third:} As in the first two panels, but H$\alpha$+[N~II] emission from MUSE in red; \textit{Fourth:} Smoothed 0.3-3~keV \textit{Chandra} image, with contours from H$\alpha$+[N~II] contours from MUSE in white.}
    \label{fig:ring_ha}
\end{figure*}

The first three panels in Fig. \ref{fig:ring_ha} show RGB composite images of the X-ray emission in the star-forming ring (0.3-1.5~keV in green, 1.5-3~keV in blue), compared to different \textit{HST} and MUSE observations. The first panel compares the soft X-ray emission with the old stellar population imaged with \textit{HST} F160W (in red); the second panel compares the soft X-ray emission with the young stellar population imaged with \textit{HST} F438W (in red); the third panel compares the soft X-ray emission with the warm 10$^{4}$~K gas traced by H$\alpha$+[N~II] and imaged with MUSE (in red). 

The X-ray emission in the star-forming ring is strongly associated with star formation, as shown in the second and third panels. However, the first panel also reveals a component of old stellar population within the ring, indicating that both young and old stellar populations contribute to the observed emission. The second panel highlights the correlation between the young stellar population - dominated by H~II regions and their OB stars - and the soft X-ray emission. This correlation, though evident, is not perfectly matched in all locations. For instance, NGC~5728~X-1 (Section \ref{sec:Two_Luminous_X-ray_Point_Sources}) appears co-located with bright regions in both the F160W and F438W images, unlike NGC~5728~X-2.

The third panel shows the spatial distribution of soft X-ray emission compared to the warm ionized gas traced by H$\alpha$+[N~II]. As expected, there is a strong correlation between the two gas phases, which is consistent with the close association between young stars and ionized gas in star-forming regions. The bright point-like source NGC~5728~X-2 is located in a region devoid of H$\alpha$+[N~II] emission, while NGC~5728~X-1 is coincident with a region of strong H$\alpha$+[N~II] emission. The fourth panel of Fig. \ref{fig:ring_ha} shows the 0.3-3~keV \textit{Chandra} image (as in Fig. \ref{fig:chandra_ring}, central panel) with H$\alpha$+[N~II] contours from MUSE overlaid in white, highlighting the correspondence between the soft X-ray emission and the warm ionized gas in the region.

\subsubsection{Physical Properties of the Hot ISM in the Ring}
\label{sec:Physical Properties of the Hot Gas in the Ring}

The results of the spectral fitting analysis in Section \ref{sec:The_Circumnuclear_Star_Forming _Ring} may be used to derive physical properties of the hot gas within the circumnuclear star-forming ring. The normalization of the thermal component is proportional to the emission measure, ${\rm EM}$\footnote{https://heasarc.gsfc.nasa.gov/xanadu/xspec/manual/XSmodelApec.html}, given by:

\begin{equation}
\label{eq:emission_measure}
    {\rm EM} = \frac{10^{-14}}{4\pi[D(1+z)]^{2}} \int n_{e}n_{\rm H}\eta dV
\end{equation}

where $n_{e}$ is the average electron density (cm$^{-3}$), $n_{\rm H}$ is the average hydrogen density (cm$^{-3}$), $dV$ is the volume element of the emitting region (cm$^{-3}$), $\eta$ is the filling factor, $D$ is the angular size distance to the source (cm), and $z$ is the redshift. 

Following \citet{baldi2006a}, we assume a depth of 200$K$ pc for the emitting volume of the ring (Fig. \ref{fig:chandra_ring}, left panel), where $K$ is a variable parameter ($K=1$ corresponds to the typical depth of a spiral disk). The total area of the emitting region is $A=1.36$~kpc$^{2}$, and the volume is $V=0.272K$ kpc$^{3}$. Using this, we calculate the electron density of the thermal gas as $n_{e}=0.14\eta^{-1/2}K^{-1/2}$~cm$^{-3}$. Thus, the total mass of hot gas within the ring is $M=7.9\times10^{5}\eta^{1/2}K^{1/2}~M_{\odot}$. These quantities depend both on the depth-variable parameter $K$ and on the filling factor $\eta$ of the gas.

Given the electron density $n_{e}$ and the best-fit temperature from the model ($kT$=0.44$\pm$0.17 keV), we estimate the following properties of the gas in the ring: pressure, calculated as $p=2n_{e}kT=2.0\times10^{-10}\eta^{-1/2}K^{-1/2}$~dyne~cm$^{-2}$, and the thermal energy, $E_{\rm th}=3n_{e}VkT=2.4\times10^{54}\eta^{1/2}K^{1/2}$~ergs. Using the model luminosity in the soft band (Section \ref{sec:The_Circumnuclear_Star_Forming _Ring}), we estimate the cooling time of the gas, defined as the ratio of the thermal energy to the X-ray luminosity of the thermal component, $\tau=E_{\rm th}/{L}^{\mathrm{model}}_{0.3-3.0\:\mathrm{keV}}=1.9\times10^{8}\eta^{1/2}K^{1/2}$~years. 

These derived physical properties are summarized in Table \ref{tab:ring} as a function of the filling factor $\eta$ and depth-variable parameter $K$. 

\begin{table*}
\centering
\caption{Physical properties of the hot ISM gas in the star-forming ring as a function of the filling factor, $\eta$, and the depth-variable parameter, $K$.}
\begin{tabular}{ccccc}
\hline
\hline
 ${n}_{e}$ & $M$ & $p$ & ${E}_{\rm th}$ & $\tau_{c}$ \\
($10^{-2}\eta^{-1/2}K^{-1/2} \; \mathrm{cm}^{-3}$) & ($10^{5}\eta^{1/2}K^{1/2} \; \mathrm{M}_{\odot}$) & ($10^{-10}\eta^{-1/2}K^{-1/2} \; \mathrm{dyne} \; \mathrm{cm}^{-2}$) & ($10^{54}\eta^{1/2}K^{1/2} \; \mathrm{ergs}$) & ($10^{6}\eta^{1/2}K^{1/2} \; \mathrm{yr}$) \\
\hline
$14.0$ & $7.9$ & $2.0$ & $2.4$ & $193.2$ \\
\hline
\label{tab:ring}
\end{tabular}
\end{table*}

\subsubsection{The Origin of the X-ray Emission in the Ring}
\label{sec:Ring Energetics}

As discussed in Section \ref{sec:The_Circumnuclear_Star_Forming _Ring}, the X-ray spectrum of the star-forming ring (Fig. \ref{fig:chandra_ring}) requires a high absorption column, $N_{H}=(6.0\pm2.1)\times10^{21}$~cm$^{-2}$, and a thermal component with $kT=0.44\pm0.17$~keV. The large value for the column density may be attributed to the presence of co-spatial CO and dust lanes, increasing the obscuration in the region. The gas temperatures in the ring are about 4-5 higher than those of the Galactic hot ISM, but are consistent with values observed in the star-forming ring of NGC 1365 ($kT=0.6\pm0.05$~keV, \citealt{wang2009a}) and the hot ISM in regions of the Antennae Galaxies (NGC 4038/4039, \citealt{baldi2006a}). Such temperatures are consistent with an origin in starburst superwinds, as in NGC 3256 \citep{moran1999a,lira2002a}, and NGC 253 \citep{pietsch2001a,strickland2000a}.

We note, however, that there are uncertainties in the model fits, the physical depth of the emitting region, and the X-ray filling factor. For instance, assuming a depth of 200~pc for the emitting region ($K=1$) and a filling factor of unity ($\eta=1$), as in NGC 1365 \citep{wang2009a} and the Antennae \citep{baldi2006a}, uncertainties in the temperature ($kT=0.44\pm0.17$~keV) introduce factors of $\sim$2.3 range in the pressure ($p$), thermal energy ($E_{\rm th}$), and cooling times ($\tau$).

To investigate whether the thermal energy in the star-forming ring could be powered by thermalization of SNe shocks, we estimate the kinetic energy input from SNe during the starburst period, and compare it to the observed thermal energy, $E_{\rm th}$. \citet{durre2018a} estimated a supernova rate (SNR) of SNR$\sim$0.4~yr$^{-1}$ for NGC 5728, using the total VLA 6cm flux. Given that the energy released in a supernova explosion is $\sim E_{\rm SN}\sim10^{51}$~ergs, and assuming that 10\% of this kinetic energy is converted into thermal energy \citep{chevalier1985a, thornton1998a}, the total energy deposited over the $\sim$6.2 Myr period of active star formation \citep{durre2018a} is $\sim E_{\rm SN}$=2.5$\times$10$^{56}$~ergs.

By comparison, the total thermal energy derived from the spectral fit (Table \ref{tab:ring}) is $E_{\rm th}$=2.4$\times10^{54}\eta^{1/2}K^{1/2}$~ergs. Thus, the observed thermal energy is $\sim$1\% of the energy supplied by SNe, assuming $\eta=1$, $K=1$. This suggests that most of the thermal energy produced by SNe has been lost, either through radiative cooling or thermal conduction, or through a wind. 

Radiative cooling of the hot gas in the ring is ruled out by the cooling times of $\tau_{c}=193$~Myr (Table \ref{tab:ring}), which is a factor of $\sim$30 times longer than the starburst period (assuming $\eta=1$ and $K=1$). A factor of 900 decrease in $\eta \times K$ would be necessary to reduce the cooling times to a comparable scale. Giving the spatial distribution of hot and warm ISM in the ring, as shown by the comparison between the X-ray and H$\alpha$+[N~II] emissions in Fig. \ref{fig:ring_ha}, it is possible the two gas phases are intermingled, leading to an X-ray filling factor $\eta<1$. This is consistent with recent 3D hydrodynamical simulation results, which suggest $0.17<\eta<0.44$ for SNR between Galactic and 16 times the Galactic value \citep[e.g.,][]{avillez2004a}. However, even for $\eta=0.2$, achieving such a large decrease in $\eta \times K$ would require a depth for the emitting region of $\sim$1 pc, which seems highly unlikely.

The conductive cooling timescale for a region of length $L$, temperature $T$, and density $n_{e}$ is:

\begin{equation}
\label{eq:conductive_cooling_time}
    t_{\rm cond}=\frac{5n_{e}kTL^{3}}{10^{-5}T^{5/2}(T/L)L^{2}}= (9\times10^{8})n_{e}T^{-5/2}_{0.5}L^{2}_{\rm pc}~ {\rm s}
\end{equation}

\noindent where $T_{0.5}$ is the temperature in units of 0.5~keV, and $L_{\rm pc}$ is the length of the emitting region in pc \citep[e.g.,][]{baldi2006a}. For $n_{e}$=0.14~cm$^{-3}$, $kT$=0.44~keV, and $L=1.5$~kpc (the average length of the emitting regions in Fig. \ref{fig:chandra_ring}), this yields a conductive timescale of $\sim$12.4 Myr, assuming full Spitzer conductivity \citep{spitzer1962a}. In practice, values between $\sim$20\% and 40\% of the Spitzer value are typical in clusters of galaxies \citep{kim2003b}, leading to a conductive timescale of $\sim$40 Myr. 

The 12.4 Myr conductive timescale is $\sim$15 times shorter than the cooling timescale of 193~Myr (assuming $\eta=1$, and $K=1$). A factor of $\sim$225 decrease in $\eta \times K$ would make the conductive cooling timescale comparable to the radiative cooling timescale, but this would imply a depth for the emitting region of $\sim$4.5 pc for $\eta$=0.2, which seems unlikely.
 
Thermal energy loss by a wind remains a possibility. The rate of energy dissipation by radiation can be estimated as:
\begin{equation}
\label{eq:energy_loss}
    \dot E_{\rm loss}=\beta n_{h}^{2}\Lambda(T)V~{\rm erg~s^{-1}}
\end{equation}

\noindent where $\beta=10$ accounts for enhanced cooling rate due to internal temperature and density variations, which also takes into account the effects of enhanced radiation from the conductive interfaces between the clouds and hot gas \citep{mckee1977a}; $\Lambda= 6.2\times10^{-19}T^{-0.6}$~ergs~cm$^{3}$~s$^{-1}$ is the cooling function for gas temperatures between $10^{5}~K<T<4\times10^{7}~K$ \citep{raymond1976a}. 

Equation \ref{eq:energy_loss} yields an energy loss rate of $\dot E_{\rm loss}=6.4\times10^{40}$~erg~s$^{-1}$, for $\eta=1$, $K=1$. Giving the total energy supplied by SNe over the period of active star formation ($E_{\rm SN}=2.5\times10^{56}$ ergs), the dissipation time via a wind would be $t_{\rm wind}\sim$124~Myr, which is $\sim1.5$ times shorter than the cooling time of 193~Myr (assuming $\eta=1$, $K=1$). However, a factor of $\sim$1.5 decrease in $\eta$ to $\eta=0.66$ would yield energy dissipation times that are comparable to the cooling times calculated for the hot gas in the ring in Table \ref{tab:ring}. 

\subsubsection{Compact Sources}
\label{sec:Compact Sources}

Of the two point-like X-ray sources detected near the star-forming ring in NGC 5728 (Section \ref{sec:Two_Luminous_X-ray_Point_Sources} and Appendix \ref{sec:appendix}), NGC~5728~X-2 does not have any obvious optical or UV counterpart, while NGC~5728~X-1 is located in a region of faint UV and H$\alpha$+[N~II] emission (Fig. \ref{fig:ring_ha}). 

The soft spectral indices ($\Gamma$=1.6-2.3) derived from the spectral analysis (Appendix \ref{sec:appendix}, Table \ref{tab:ps}) suggest that these sources may be X-ray binaries. Their luminosities range between 5.0-8.2$\times10^{38}$~erg~s$^{-1}$, which exceed the Eddington limit for a neutron star (L$_{X}\sim3\times10^{38}$~erg~s$^{-1}$; see e.g. \citealt{king2001a}). If these sources are neutron stars binaries, they would likely be in a super-Eddington accretion state. 

Another possibility is that these sources are accreting black hole binaries. Accreting black holes typically exhibit two primary spectral states: the high/soft state and the low/hard state. The high/soft state occurs at high accretion rates ($\dot M_{\rm Edd}\sim0.1-0.5$ of the Eddington accretion rate). In this state, the X-ray spectrum is dominated by thermal blackbody emission from the accretion disk, with typical temperatures of $kT\sim$1~keV, and the power-law component associated with the corona has $\Gamma \sim$2-3 \citep[e.g.,][]{gilfanov2004a, fornasini2024a}. Although the present data is limited, if these sources are black hole binaries the fitting results in Table \ref{tab:ps} suggest that NGC~5728~X-2 could be a disk-dominated, high-state black hole HMXB. For NGC~5728~X-1, the best-fit models yield similarly small $\chi^{2}_{\nu}$ (Table \ref{tab:ps}), but the model parameters remain unconstrained.

To further investigate the nature of these sources, we use the universal HMXB luminosity function (LF) proposed by \citet{grimm2003a}, which normalizes the number of sources to the star-formation rate (SFR) using the following equation:

\begin{equation}
\label{eq:luminosity_function}
    N(>L)=5.4~{\rm SFR} ((L/10^{38} ) ^{-0.61}- 210^{-0.61})
\end{equation}

This equation describes a differential LF with a slope of $\alpha$=1.6 and a cut-off luminosity of $L_{\rm X_{(2-10~{\rm keV})}}=2\times10^{40}$~erg~s$^{-1}$. For a star-formation rate of SFR=3.5~$M_{\odot}$~yr$^{-1}$ \citep{durre2018a}, the estimated number of sources with $L>7\times10^{38}$~erg~s$^{-1}$ is $\sim5$. 

Alternatively, following another approach suggested by \citet{grimm2003a}, we can relate the SFR to the total X-ray luminosity from HMXBs. Using this method, we derive a lower SFR of $\sim$0.75~$M_{\odot}$~yr$^{-1}$, which estimates $N(>L=7\times10^{38}$~erg~s$^{-1}$)$\sim1-2$. 

In both cases, detecting two X-ray sources in the star-forming ring is consistent with the HMXB luminosity function, given the statistical uncertainty. The observed number of sources falls within the expected range, supporting the interpretation that these compact sources are HMXBs related to the ongoing star formation in NGC 5728.

\section{Conclusions}
\label{sec:conclusions}

In this paper we have conducted a detailed spatial and spectral analysis of the diffuse soft emission in the nearby Seyfert 2 galaxy NGC 5728, examining the cumulative 260 ks exposure obtained with \textit{Chandra} ACIS-S. We compared the soft X-ray emission with the distribution and properties of the cold and warm ISM in this galaxy. Our main findings are summarized as follows:

1. The soft X-ray emission extends out to $r\sim$1.4 kpc within the SE bicone, while the NW bicone shows a shorter extent ($r\sim$1 kpc), likely due to obscuration of this region by the nuclear star-forming disk. 

2. Hardness ratio images and spectral fitting indicate increased X-ray absorption perpendicular to the bicone axis (in region R2). This absorption follows a similar position angle to that of the dusty spirals seen in the \textit{HST} (V-H) color map and the warped inner molecular disk mapped in CO (2-1). These features likely trace cold material feeding the central SMBH \citep{shimizu2019a}.

3. We detected diffuse X-ray emission from the star-forming ring, located at $r\sim$1~kpc from the AGN. The thermal component of the ring has a temperature of $kT=$0.44~keV, consistent with starburst-driven superwinds \citep[e.g.,][]{lira2002a,strickland2000a}, and comparable to those found in similar AGNs, such as NGC 1365 \citep{wang2009a}. The physical parameters of the hot gas in the ring suggest cooling times on the order of $\tau_{c}=$193$\eta^{1/2}K^{1/2}$~Myr, and a total mass of $M=$7.9$\times10^{5}\eta^{1/2}K^{1/2}$M$_{\odot}$. The thermal energy content of this gas is $\sim$1\% of the energy expected from supernovae, suggesting that much of the energy may have been dissipated through non-radiative processes, most likely stellar winds. 

4. Two isolated X-ray point-like sources were detected near the star-forming ring, within the inner 1 kpc. Their soft spectra are consistent with X-ray binaries, with X-ray luminosities in the range $L_{X}=5.0-8.2\times10^{38}$~erg~s$^{-1}$. These detections are consistent with a population of HMXBs in an actively star forming galaxy.

\bibliographystyle{aasjournal}
\bibliography{anna_bibliography}

\appendix
\section{Two Luminous X-ray Point Sources}
\label{sec:appendix}

Two bright X-ray point-like sources, located near the star-forming ring (Figs. \ref{fig:soft_morphology_rgb} and \ref{fig:chandra_ring}), were detected primarily in the soft ($<$3~keV) X-ray band \citepalias[see also][]{trindadefalcao2023a}.

For sufficiently bright sources, the X-ray spectrum can be used to constrain the nature of the emission. X-ray Binaries (XRBs) and AGNs typically exhibit power-law spectra. If these sources are XRBs, their photon index can be used to constrain the nature of the compact object: neutron stars (NS) usually have harder X-ray spectra with  $\Gamma<$1.5, while black holes (BHs) exhibit softer spectra with $\Gamma\sim$1.5-2.0 due to the absence of a solid surface \citep[e.g.,][]{mcclintock2006a}. In contrast, foreground stars and supernova remnants (SNRs) display significantly softer X-ray spectra, with emission lines suggestive of a thermal plasma.

\begin{figure}[h!]
    \centering
    \includegraphics[width=\textwidth]{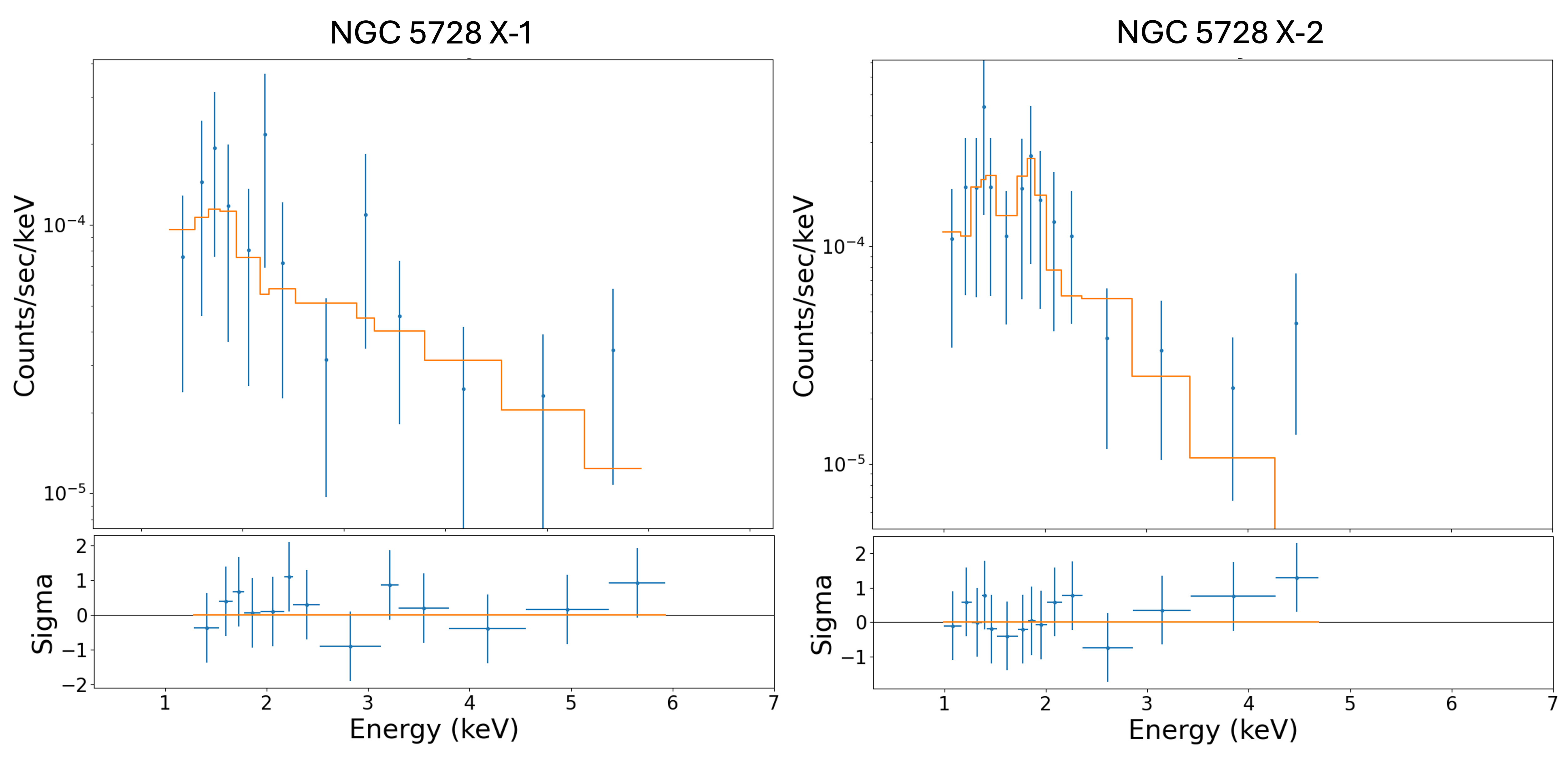}
    \caption{X-ray spectra of the two bright point sources detected near the star-forming ring in the \textit{Chandra} FOV of NGC 5728. Best-fit power-law model is shown for NGC~5728~X-1, and best-fit thermal model is shown for NGC~5728~X-2. Residuals are shown on the bottom panels.}
    \label{fig:ps_spec}
\end{figure}

We used \texttt{CIAO} \textit{specextract} to extract the spectra of each X-ray source using a 0.8$''$ circular region (Fig. \ref{fig:ps_spec}, binned at 5 counts per bin). Initially, we fit each dataset with an absorbed thermal model (\texttt{xsapec}, see Section \ref{sec:The_Circumnuclear_Star_Forming _Ring}), assuming 1$\times$solar abundances. The rationale for using a thermal model is that the excess of soft X-ray emission may originate from collisionally-ionized gas, possibly from SNRs. The absorption column density is left free to vary above the Galactic value $N_{H_{\rm galactic}}=7.53\times10^{20}$~cm$^{-2}$ during the fit (see \citetalias{trindadefalcao2023a}).

For NGC~5728~X-1 (north of the AGN, Fig. \ref{fig:chandra_ring}), the thermal model fitting returns unconstrained values for the gas temperature and an absorbing column of $N_{H}^{\rm thermal}<10^{22}$~cm$^{-2}$, with $\chi^{2}_{\nu}=0.47$. For NGC~5728~X-2 (northeast of the AGN, Fig. \ref{fig:chandra_ring}), the best-fit temperature was $kT=$1.07$\pm$0.31~keV, with an absorption column density of $N_{H}^{\rm thermal}=(1.41\pm0.40)\times10^{22}$~cm$^{-2}$, and $\chi^{2}_{\nu}=0.42$.

\smallskip
We also explored two additional spectral models: 

(1) Absorbed power-law model: The power-law model was used to account for the possibility of excess of soft X-ray emission arising from a low-state black hole binary, which typically exhibits a non-thermal spectrum \citep{mcclintock2006a}.

\texttt{xstbabs.abs\_soft$\times$xspowerlaw.po};

(2) Absorbed accretion disk model: The accretion disk model was used to test the hypothesis that the soft X-ray emission in the spectra originates from a high-state black hole binary, which can be modeled with a multi-temperature accretion disk (\texttt{xsdiskbb}).

\texttt{xstbabs.abs\_soft$\times$xsdiskbb\footnote{https://cxc.cfa.harvard.edu/sherpa/ahelp/xsdiskbb.html}.disk}. 

For both models, the total column density was left free to vary above the Galactic value, and spectral fitting was performed over the 0.3-7~keV range.

\smallskip
NGC~5728~X-1: The absorbed power-law model for NGC~5728~X-1 provided unconstrained values for the absorbing column, with a photon index of $\Gamma=1.6\pm1.3$, but no significant improvement in $\chi^{2}_{\nu}$ compared to the thermal model (Table \ref{tab:ps}). Similarly, the accretion disk model yielded unconstrained values for both the column density and the disk temperature. 

\smallskip
NGC~5728~X-2: For NGC~5728~X-2, the power-law model provided a slightly better fit ($\chi^{2}_{\nu}=0.34$) than the thermal model, though the column density was unconstrained and the photon index was steep, but poorly constrained ($\Gamma=2.3\pm1.2$). The accretion disk model returned a temperature at the inner disk radius of $kT=$1.1$\pm$0.6~keV and $\chi^{2}_{\nu}=0.36$, but again, with an unconstrained absorbing column.

\smallskip 
Despite the low signal-to-noise ratio (S/N) of the data (Fig. \ref{fig:ps_spec}), the similar $\chi^{2}_{\nu}$ values across different models suggest that all models provide acceptable fits (Table \ref{tab:ps}). The estimated X-ray luminosities are $L^{\rm model}_{0.3-7~{\rm keV}}=7.8-8.2\times10^{38}$~erg~s$^{-1}$ for NGC~5728~X-1, and $L^{\rm model}_{0.3-7~{\rm keV}}=5.0-7.4\times10^{38}$~erg~s$^{-1}$ for NGC~5728~X-2 (Table \ref{tab:ps}). These luminosities exceed the Eddington limit for a $\sim$1 solar mass accreting object ($L\sim$1.3$\times$10$^{38}$~erg~s$^{-1}$), but fall within the canonical range for HMXBs \citep[e.g.,][]{gilfanov2004a, fabbiano2006a}, which likely contribute significantly to the observed luminosities. 

\begin{table}[h!]
\footnotesize
\centering
\caption{Details for the two X-ray point-like sources.}
\label{tab:ps}
\begin{tabular}[t]{c c c c c c c c c c c}
\hline 
\hline
Source& Position &Net Counts& Fitting& $N_{\rm H}$&$\Gamma$& $kT$& ${\rm d.o.f.}$&$\chi^{2}_{\nu}$& $f^{\rm model}_{0.3-7~{\rm keV}}$&$L^{\rm model}_{0.3-7~{\rm keV}}$\\ 
& (J2000.0)&(0.3-7~keV)	& Model$^a$	& (10$^{22}$cm$^{-2}$)	&	& (keV)	&	&	& (erg~s$^{-1}$~cm$^{-2}$)&(erg~s$^{-1}$) \\
\hline
NGC~5728~X-1& R.A.:$14^{\rm h}42^{\rm m}23^{\rm s}.9$& 77$\pm$9 & thermal&$<10^{22}$& -&unconstrained &10&0.47&4.1$\times10^{-15}$&8.2$\times10^{38}$\\

&Dec.:$-17\degree15'05''3$ & &power-law&unconstrained&1.6$\pm$1.3& -&10&0.47&3.9$\times10^{-15}$&7.8$\times10^{38}$\\

& && disk&unconstrained& -&unconstrained &10&0.49&4.1$\times10^{-15}$&8.2$\times10^{38}$\\

\hline

NGC~5728~X-2&R.A.:$14^{\rm h}42^{\rm m}24^{\rm s}.3$& 84$\pm$9 & thermal&1.4$\pm$0.4& -&1.1$\pm$0.3&12&0.42&2.5$\times10^{-15}$&5.0$\times10^{38}$\\ 

&Dec.:$-17\degree15'08''8$ &&power-law&unconstrained&2.3$\pm$1.2& -&12&0.34&3.7$\times10^{-15}$&7.4$\times10^{38}$\\

& && disk&unconstrained& -&1.1$\pm$0.6 &12&0.36&3.7$\times10^{-15}$&7.4$\times10^{38}$\\

\hline

\end{tabular}
\end{table}

\end{document}